\documentclass[acmtog]{acmart}

\AtBeginDocument{%
  \providecommand\BibTeX{{%
    \normalfont B\kern-0.5em{\scshape i\kern-0.25em b}\kern-0.8em\TeX}}}

\copyrightyear{2025}
\acmYear{2025}
\setcopyright{cc}
\setcctype{by}
\acmConference[SA Conference Papers '25]{SIGGRAPH Asia 2025 Conference Papers}{December 15--18, 2025}{Hong Kong, Hong Kong}
\acmBooktitle{SIGGRAPH Asia 2025 Conference Papers (SA Conference Papers '25), December 15--18, 2025, Hong Kong, Hong Kong}\acmDOI{10.1145/3757377.3763840}
\acmISBN{979-8-4007-2137-3/2025/12}

\usepackage[boxed,ruled,linesnumbered]{algorithm2e} 
\let\oldnl\nl
\newcommand{\nonl}{\renewcommand{\nl}{\let\nl\oldnl}}

\SetAlFnt{\small}
\SetAlCapFnt{\small}
\SetAlCapNameFnt{\small}
\SetAlCapHSkip{0pt}
\IncMargin{-\parindent}
\newlength\savedwidth

\usepackage{enumitem} 
\usepackage{color, colortbl}

\DeclareMathOperator*{\argmin}{\arg\!\min}
\usepackage{soul}
\usepackage{multirow}
\usepackage{bbding}
\usepackage{pdfpages}
\theoremstyle{definition}

\newcommand{\violet}[1]{\textcolor{violet}{#1}}

\newcommand{\model}{\mathcal{S}}
\newcommand{\bmodel}{\partial \mathbf{S}}

\newcommand{\ma}{\mathcal{M}}
\newcommand{\mmesh}{\mathcal{M}_s}

\newcommand{\msphere}{\mathbf{m}}
\newcommand{\mcenter}{\boldsymbol \theta}
\newcommand{\medge}{\mathbf{e}}
\newcommand{\mface}{\mathbf{f}}
\newcommand{\anypoint}{\mathbf{x}}
\newcommand{\anypointproj}{\mathbf{p}}
\newcommand{\anypointnormal}{\mathbf{n}}

\newcommand{\singular}{\sigma}

\newcommand{\kernel}{\sigma}

\newcommand{\rpc}{\mathbf{\omega_c}} 

\newcommand{\tanpoint}{\mathbf{p}}

\newcommand{\energy}{E}

\newcommand{\ie}{\textit{i.e., }}
\newcommand{\eg}{\textit{e.g., }}
\newcommand{\hd}{\textit{HD}}
\newcommand{\triangleQ}{\textit{TQ}}

\newcommand{\supmaterial}[1]{{\color{blue}#1}}

\begin{document}
\title{MATStruct: High-Quality Medial Mesh Computation via Structure-aware Variational Optimization}


\author{Ningna Wang}
\affiliation{%
  \institution{University of Texas at Dallas}
  \state{Texas}
  \country{USA}
}
\email{ningna.wang@utdallas.edu}

\author{Rui Xu}
\affiliation{%
  \institution{University of Hong Kong}
  \city{Hong Kong}
  \country{China}
}
\email{ruixu1999@connect.hku.hk}

\author{Yibo Yin}
\affiliation{%
  \institution{Wuhan University}
  \city{Wuhan}
  \country{China}
}
\email{yinyb811@gmail.com}

\author{Zichun Zhong}
\affiliation{%
  \institution{Wayne State University}
  \city{Detroit}
  \country{USA}
}
\email{zichunzhong@wayne.edu}

\author{Taku Komura}
\affiliation{%
  \institution{University of Hong Kong}
  \city{Hong Kong}
  \country{China}
}
\email{taku@cs.hku.hk}

\author{Wenping Wang}
\affiliation{%
  \institution{Texas A\&M University}
  \state{Texas}
  \country{USA}
}
\email{wenping@tamu.edu}

\author{Xiaohu Guo}\authornote{Corresponding author}
\affiliation{
  \institution{University of Texas at Dallas}
  \state{Texas}
  \country{USA}
}
\email{xguo@utdallas.edu}

\renewcommand\shortauthors{Wang et. al}

\begin{abstract}
We propose a novel optimization framework for computing the medial axis transform that simultaneously preserves the medial structure and ensures high medial mesh quality. The \textit{medial structure}, consisting of interconnected \textit{sheets}, \textit{seams}, and \textit{junctions}, provides a natural volumetric decomposition of a 3D shape. Our method introduces a structure-aware, particle-based optimization pipeline guided by the restricted power diagram (RPD), which partitions the input volume into convex cells whose dual encodes the connectivity of the medial mesh. Structure-awareness is enforced through a spherical quadratic error metric (SQEM) projection that constrains the movement of medial spheres, while a Gaussian kernel energy encourages an even spatial distribution. Compared to feature-preserving methods such as MATFP~\cite{2022MATFP} and MATTopo~\cite{wang2024mattopo}, our approach produces cleaner medial structures with significantly improved mesh quality. In contrast to voxel-based, point-cloud-based, and variational methods, our framework is the first to integrate structural awareness into the optimization process, yielding medial meshes with explicit structural decomposition, topological correctness, and geometric fidelity. Our \href{https://github.com/ningnawang/MATStruct}{\violet{code}} is available at \href{https://ningnawang.github.io/projects/2025_matstruct/}{\violet{our project website}}.
\end{abstract}

%


%

\begin{teaserfigure}
    \centering
    \includegraphics[width=\linewidth]{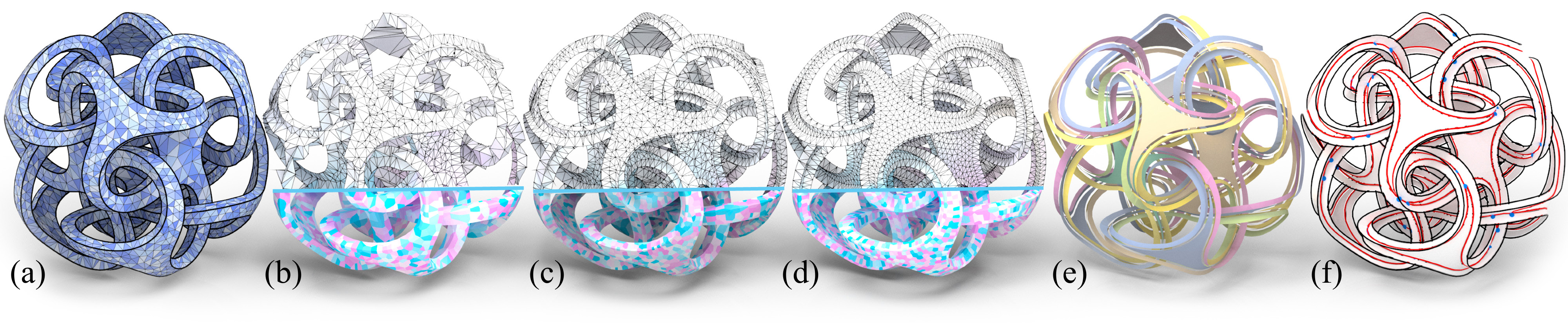}
    \vspace{-20pt}
    \caption{The pipeline of our structure-aware medial axis transform framework. (a) Starting from a tetrahedral mesh with pre-detected convex sharp edges and corners (black curves), our method iteratively optimizes the medial mesh (b)(c)(d) with structure-awareness. The resulting medial structure includes  (e) medial sheets (slightly shrunk inward and rendered in randomly assigned colors) and (f) seams shown in red, and medial junctions shown as blue spheres. Our method is the first to produce clean medial structure representations (e)(f) compared to existing approaches. 
    }
    \label{fig:teaser}
\end{teaserfigure}

\maketitle

\section{Introduction}


The medial axis~\cite{blum1967transformation} is a fundamental geometric structure that captures both the topological and geometric properties of a shape. Defined as the set of points within a shape $\model$ that have two or more nearest neighbors on the boundary $\bmodel$, the medial axis $\ma$ effectively represents the skeleton of the shape. The \textit{medial axis transform} (MAT) further augments the medial axis with a radius function, creating a concise representation that encodes the geometry and topology of the input shape. Given the computational complexity of deriving an exact 3D medial axis, existing approaches typically resort to approximations that aim to preserve key properties such as centeredness, topological equivalence, and geometric reconstructability~\cite{tagliasacchi20163d}.

\begin{figure}[!h]
    \centering
    \includegraphics[width=0.9\linewidth]{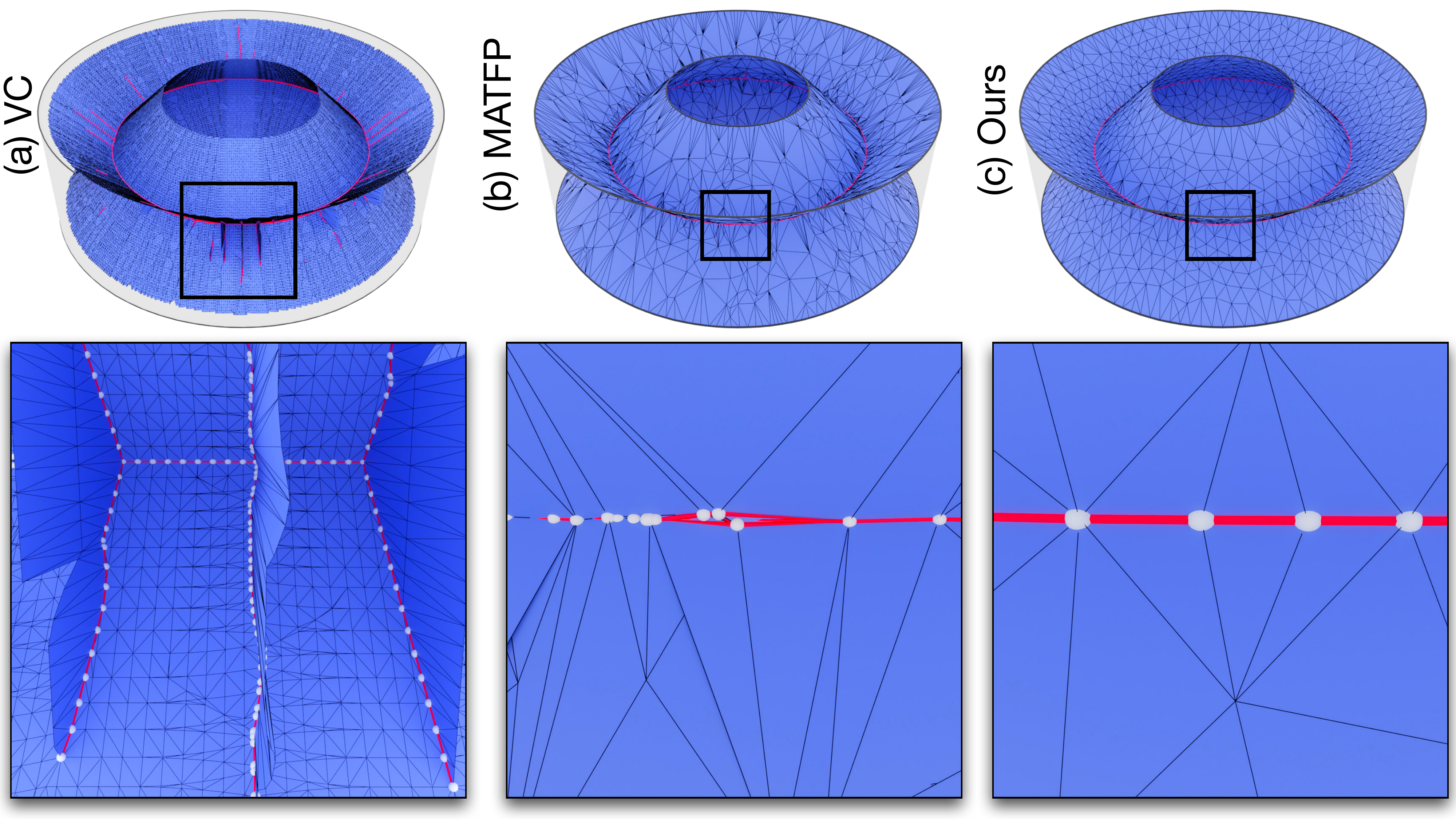}
    \caption{The medial mesh generated for a hollow cylinder (rendered transparently) using VC~\cite{yan2018voxel}, MATFP~\cite{2022MATFP}, and our method. Medial seams are shown as red lines, with corresponding spheres highlighted. (a) VC produces excessive medial sheets, resulting in redundant and spurious seams. (b) MATFP generates densely clustered medial spheres around seams, leading to numerical instability and erroneous seam connections (see Sec.~\ref{sec:pre_challenge} for details). (c) Our method produces a high-quality 3D medial mesh with clear structural decomposition. Note that we show zoom-ins at two different scales to better reveal the redundant sheets in VC.}
    \label{fig:intro_why}
\end{figure}

The \textit{medial structure}, composed of interconnected \textit{sheets} (Fig.~\ref{fig:teaser} (e)), \textit{seams}, and \textit{junctions} (Fig.~\ref{fig:teaser} (f))~\cite{giblin2004formal}, provides a natural volumetric decomposition of the medial axis. This structure is particularly valuable for downstream applications such as shape analysis, recognition, and matching, as it remains largely stable under small shape deformations and preserves the topological relationships among key medial points~\cite{leymarie2007medial, chang2008regularizing}.
Classical methods such as PC~\cite{amenta2001power} and SAT~\cite{miklos2010sat} neglect the underlying medial structure and fail to maintain the thinness property of the medial axis. Their approximated medial meshes often contain many 3-dimensional cells (`closed pockets'), making it impossible to extract the medial structure from their results.
The seminal work VoxelCore (VC)~\cite{yan2018voxel} constructs a subset of the Voronoi diagram from boundary vertices, enabling medial structure identification via non-manifold analysis~\cite{chang2008medial}. However, it often produces excessive and redundant medial sheets that are difficult to prune effectively, as shown in Fig.~\ref{fig:intro_why}~(a).
Recent methods such as MATFP~\cite{2022MATFP} and MATTopo~\cite{wang2024mattopo} leverage restricted power diagrams (RPD) to approximate the medial axis transform (MAT), demonstrating improved preservation of medial structures, especially for CAD models. Nevertheless, both methods share a common limitation: they insert medial spheres on seams upon detecting deficiencies but do not update their positions during optimization. This leads to numerical instability and degraded structural clarity, as illustrated in Fig.~\ref{fig:intro_why} (b). Moreover, both rely on surface-based RPD classification, which frequently misclassifies medial spheres in regions with dense or complex topology. As a result, the final medial meshes often suffer from poor triangle quality, undermining their geometric fidelity and structural soundness.

In this work, we introduce a novel optimization framework for computing the medial axis transform (MAT) that jointly preserves medial structure and achieves high-quality medial mesh generation. Our approach combines a particle-based optimization scheme with a restricted power diagram (RPD) based spherical quadratic error metric (SQEM). The particle-based scheme incorporates a Gaussian energy that promotes an even spatial distribution of medial spheres, while the SQEM constrains their movement to ensure structure-aware optimization. Together, these components enable accurate extraction of medial structures with clean connectivity and well-shaped mesh elements. 
To prevent medial spheres from drifting away from the underlying medial axis, our SQEM formulation imposes directional constraints derived from the RPD, ensuring that spheres remain within structurally valid regions throughout the optimization.

While \textit{Centroidal Voronoi Tessellation} (CVT)~\cite{du1999centroidal} offers an alternative for optimizing point distributions in volumetric domains, it requires an additional volume decomposition via a restricted Voronoi diagram (RVD), separate from the existing RPD structure used in our framework. Moreover, CVT aims to distribute points uniformly across the entire shape volume, which conflicts with our objective of concentrating medial spheres specifically along the medial structure. In contrast, our particle-based approach does not require auxiliary volume decomposition and integrates seamlessly with the RPD-based shape representation. 
Methods such as \textit{Optimal Delaunay Triangulations} (ODT) and \textit{Laplacian smoothing}~\cite{chen2011efficient} rely on a fixed connectivity graph (in our case, the dual of the RPD) during optimization, which is unreliable in early stages due to uneven and unstructured sphere distribution. This makes them unsuitable in our setting, as they make it difficult to enforce uniform distribution of spheres over the medial axis. 

The main contributions of this work are as follows:
\begin{itemize}
\item We present a comprehensive RPD-based optimization framework for generating high-quality 3D medial meshes with structure awareness, effectively preserving medial features, particularly in CAD models which tend to exhibit well-defined and clean geometric features characteristic of human-made designs. 

\item We propose a robust volumetric RPD-based sphere classification strategy that significantly improves the accuracy of medial sphere classification compared to surface RPD-based methods. 
\item We introduce the medial structure error ratio (MSER), a new quantitative metric for evaluating the accuracy and consistency of medial structure extraction across different methods. 
\end{itemize}





\section{Related Works}
\label{sec:related_works}

The topological and geometric properties of medial axis allow it to become the foundation for other skeletal shape descriptors~\cite{tagliasacchi20163d} and has been used in approximating~\cite{hu2022immat, hu2023s3ds, ge2023point2mm, yang2020p2mat, yang2018dmat}, simplifying~\cite{li2015qmat, dou2021coverage, yan2016erosion, wang2024coverage}, and analyzing shapes~\cite{Hu2019MATNet, lin2021point2skeleton, xu2024cwf, DroneUpdate25}. Some literature~\cite{tagliasacchi20163d, kustra2013, kustra2015} refers to the term `medial axis' as 2D skeletons and uses `medial surface' for 3D structures. For clarification, we consider `medial axis' as a broader definition that includes both 2D and 3D. In this paper, we focus exclusively on the 3D medial axis. 

\paragraph{Medial Spheres Classification}
Giblin and Kimia~\shortcite{giblin2004formal} introduced the concepts of \textit{medial sheets}, \textit{medial seams} (so called \textit{internal features}), and \textit{medial junctions}, and provided a formal classification of medial spheres based on their order of contact for organic shapes
which is also intensively discussed in seminal works~\cite{kustra2015, tagliasacchi20163d}.
Wang et al.~\shortcite{2022MATFP} extended these ideas to handle non-smooth regions, such as convex sharp edges and corners (so called \textit{external features}) that are commonly found in CAD models. 
MATFP~\cite{2022MATFP} also proposes a surface RPD-based classification strategy that groups medial spheres according to the connected components (CCs) of the surface regions within each restricted power cell (RPC). However, this surface-based approach becomes unstable in regions where multiple medial spheres are densely packed. In such cases, the spheres compete for surface CCs in overlapping regions, often leading to misclassification. To address this limitation, we propose a more robust volumetric RPD-based classification method, detailed in Sec.~\ref{sec:pre_challenge}.

\paragraph{Hierarchical Organization of the Medial Axis}
Leymarie and Kimia \shortcite{leymarie2007medial} introduced the concept of the \textit{medial scaffold}, a hierarchical graph structure composed of special medial curves (\ie medial seams) connecting special medial points (\ie medial junctions). They proposed a propagation-based algorithm to extract the scaffold from dense point clouds; however, identifying the initial sources of flow remains a major computational bottleneck.
Chang et al.~\shortcite{chang2008regularizing, chang2008medial} recovered the medial scaffold from the Voronoi diagram and focused on regularizing medial structures to ensure that similar 3D shapes yield consistent medial axes. Since VC~\cite{yan2018voxel} generates a medial mesh from a subset of the Voronoi diagram of boundary voxel samples while preserving the thinness property of the medial axis. We adopt VC as a representative baseline for comparison. In VC, the medial structure is extracted through non-manifold analysis, where seams are defined as intersections of multiple manifold sheets, and junctions as intersections of multiple seams.
Kustra et al.~\shortcite{kustra2015} apply the sphere-shrinking algorithm~\cite{ma20123shrink} to generate medial spheres from dense point clouds, slightly increasing sphere radii to approximate medial seams. However, their method cannot preserve internal medial features such as seams and junctions. This limitation stems from the fact that the sphere-shrinking algorithm only produces $T_2$ spheres—spheres tangent to exactly two surface points—thus capturing only medial sheets. It fails to generate higher-order medial spheres with three or more tangents (\eg $T_3$ on seams or $T_4$ on junctions), making accurate extraction of complex medial structures infeasible.
Recent works MATFP~\cite{2022MATFP} and MATTopo~\cite{wang2024mattopo} attempt to preserve medial seams by inserting internal feature spheres upon detecting structural deficiencies. However, their results often contain densely clustered medial spheres around seams and junctions, leading to numerical instability and the extraction of redundant seams. We refer the reader to Fig.~\ref{fig:intro_why} and Fig.~\ref{fig:pre_structure} for illustration, and to Sec.~\ref{sec:exp} for detailed experimental comparisons.

\paragraph{Medial Axis Approximation}
In addition to \textit{algebraic methods}~\cite{culver2004exact, milenkovic1993robust, sherbrooke1996algorithm} and \textit{voxel-based methods}~\cite{saha2016survey, sobiecki2014comparison, yan2018voxel, hesselink2008euclidean, rumpf2002continuous, siddiqi2002hamilton, jalba2015unified}, \textit{Voronoi-based} methods remain the most widely used for approximating the 3D medial axis. These approaches sample points on the shape boundary and extract a stable and meaningful subset of the Voronoi diagram. Examples include \textit{angle-based filtering methods}~\cite{amenta2001power, brandt1992continuous, dey2002approximate, dey2004approximating, foskey2003efficient, sud2005homotopy}, which retain medial spheres based on the angle between their closest boundary points, and $\lambda$-medial axis methods~\cite{chazal2005lambda, chazal2008smooth}, which apply a filtering threshold based on radius.
The scale axis transform (SAT)~\cite{giesen2009scale, miklos2010sat} and \textit{sphere-shrinking-based} (SS) methods apply multiplicative scaling to medial spheres, removing unstable spikes while preserving small features. MATFP~\cite{2022MATFP} initializes medial spheres from interior Voronoi vertices and refines them through iterative updates. However, these approaches often fail to simultaneously preserve the topology of the original shape and capture features at multiple scales. MATTopo~\cite{wang2024mattopo} addresses topological preservation using volumetric RPD, a strategy we also adopt to maintain homotopy equivalence.

Additionally, a range of skeletonization methods have been proposed to generate sparse skeletal representations~\cite{dou2021coverage, wang2024coverage, li2015qmat, yan2016erosion}, typically requiring either an initial medial axis approximation or a set of candidate inner balls as input. The recent variational approach VMAS~\cite{vmas2024} belongs to this category, incrementally adding medial spheres in a coarse-to-fine manner by minimizing a hybrid metric that incorporates both plane-sphere and point-sphere distances. However, VMAS does not account for feature preservation—neither external features, such as convex sharp edges from CAD models, nor internal features like seams and junctions. Furthermore, VMAS becomes unstable and failure to terminate when the number of medial spheres exceeds a few hundreds. In such cases, spheres may oscillate between configurations, causing the system to repeatedly insert and delete spheres without convergence. 
Recent deep learning methods have explored skeleton and medial axis extraction from point clouds or meshes~\cite{clemot2023neural, ge2023point2mm, lin2021point2skeleton, yang2020p2mat, kong2024quasi}. While effective on clean data, these approaches often lack guarantees on topology and structure. 


\section{Problem Statement}
\label{sec:pre}

In this section, we first define the concept of \textit{medial structure} (Sec.~\ref{sec:pre_structure}), then present two key challenges: medial sphere classification and medial sphere overcrowding, along with our proposed solutions (Sec.~\ref{sec:pre_challenge}). Our guiding principle for structuring the paper is to keep all novel contributions in the main paper, while moving content primarily related to prior work into the supplementary material. References to sections, figures, and tables from the \supmaterial{Supplementary Material} are highlighted in \supmaterial{blue} for clarity.

\subsection{Medial Structure}
\label{sec:pre_structure}

The medial axis structure, referred to as the \textit{medial structure}, is composed of connected \textit{sheets}, \textit{seams}, and \textit{junctions}~\cite{giblin2004formal}. We adopt the same classification of medial spheres as MATFP~\cite{2022MATFP}, which extends Giblin and Kimia's~\shortcite{giblin2004formal} observations to account for sharp edges and corners in CAD models. 
For a non-smooth model (\ie CAD model) that contains sharp edges and corners, either convex or concave, we use a dihedral angle less than $\pi-\phi$ and greater than $\pi+\phi$ to define the \textit{convex sharp edge} and \textit{concave sharp edge} respectively \cite{abdelkader2020vorocrust, 2022MATFP}. Here $\phi$ is a user-defined variable, and the user can also mark sharp edges manually.

\begin{figure}
	\centering
	\includegraphics[width=\linewidth]{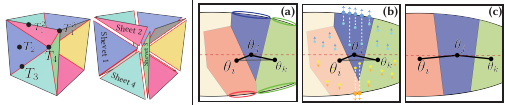}
	\vspace{-10pt}
	\caption{Left: The medial axis of a cube illustrating the medial structure described in Sec.~\ref{sec:pre_structure}.  Right: 2D illustration of classification (a), overcrowding issues (b) described in Sec.~\ref{sec:pre_challenge} using three spheres with centers $\mcenter_i$, $\mcenter_j$, and $\mcenter_k$, where two of them slightly deviate from the medial axis (shown as a red dotted line). }
	\label{fig:pre_structure}
	\vspace{-10pt}
\end{figure}

 Consider the medial axis of a cube as an example, see Fig.~\ref{fig:pre_structure} left. The interior of a cube's medial sheet comprises $T_2$ spheres. Each sheet is bounded by seams (\ie $T_3$) or external feature boundaries (\ie $T_1^2$). Seams terminate at junctions (\ie $T_4$), convex corners (\ie $T_1^3$). This classification facilitates the decomposition of the computed medial axis into distinct components. The seams and junctions define what we refer to as \textit{internal features} of the medial axis, capturing critical structural transitions within the shape. And the convex sharp edges and corners of the object define the \textit{external features}, which encode key geometric details of the original shape boundary.

We use the \textit{medial mesh} $\mmesh$~\cite{li2015qmat} to represent the approximation of a 3D medial axis $\ma$. To construct the medial mesh $\mmesh$, we adopt the same volumetric-RPD-based strategy from MATTopo~\cite{wang2024mattopo}. More details about the definition of medial mesh, the volumetric RPD, and their duality can be found in \supmaterial{Sec.~1} (Supplementary Material).

\subsection{Challenges and Our Solutions}
\label{sec:pre_challenge}

Computing medial structures is difficult because internal features like seams and junctions are hidden within the shape. Prior feature-preserving methods such as MATFP~\cite{2022MATFP} and MATTopo~\cite{wang2024mattopo} face two key challenges: (1) inaccurate medial sphere classification due to surface-only reasoning, and (2) overcrowded feature spheres near internal structures, leading to tangled connectivity.

\subsubsection{Medial Spheres Classification}

Identifying tangency contacts for medial spheres, particularly for spheres with more than two contacts like $T_3$ and $T_4$, is challenging. MATFP~\cite{2022MATFP} proposed classifying spheres via surface restricted power diagrams (surface-RPD), by counting surface-connected components of each sphere’s power cell. However, this approach assumes spheres lie exactly on the medial axis. In practice, sphere centers often deviate from the axis, as shown in Fig.~\ref{fig:pre_structure} right (a), where surface-connected regions are highlighted with ellipses. Such deviations shift the restricted power cells (RPCs) and lead to misclassifications; for instance, only $\msphere_k$ with center $\mcenter_k$ is correctly classified as a $T_2$ sphere.

Fortunately, the volumetric-RPD offers us a better tool to solve the medial sphere classification problem (Challenge 1) by providing volumetric information rather than just surface regions. 
The medial axis serves as a volumetric encoding that decomposes the shape into sub-volumes in the local region. For instance, in Fig.~\ref{fig:pre_structure} right (b), the medial axis, represented by the red dotted line, divides the shape into two local sub-volumes: an upper and a lower sub-volume. Even if the medial spheres $\msphere_i$ deviate from the medial axis, its volumetric RPCs still maintain intersections with both the upper and lower sub-volumes. Each sample within an RPC can determine its closest projection on the surface. By clustering these projections, we can accurately classify the corresponding medial sphere. 

\subsubsection{Medial Sphere Overcrowding}

To preserve internal features such as seams and junctions, prior methods like MATFP~\cite{2022MATFP} and MATTopo~\cite{wang2024mattopo} repeatedly insert feature spheres upon detecting deficiencies, but do not update their positions afterward. This often leads to densely clustered spheres near internal features, introducing two key issues. First, classification becomes more difficult, as the associated restricted power cells (RPCs) of tightly packed spheres compete for surface regions. Second, it results in excessive seam connections, creating a tangled internal structure. As shown in Fig.~\ref{fig:pre_structure} right (a), a redundant medial edge is formed between spheres $\mcenter_i$ and $\mcenter_k$.

We address this issue by optimizing the positions of inserted spheres using a structure-aware, particle-based strategy. As shown in Fig.~\ref{fig:pre_structure} right (c), after optimization, the spheres are more evenly distributed: the RPC of $\msphere_j$ now separates those of $\msphere_i$ and $\msphere_k$, leading to cleaner, more meaningful connectivity and improved triangle quality in the final medial mesh. Note that while spurious connections like the one in Fig.~\ref{fig:pre_structure} right (a) may still occur after optimization, they are significantly reduced. To further prune such artifacts, we introduce a volumetric-RPD based method for classifying and filtering medial edges and medial faces as part of a post-processing step (see Sec.~\ref{sec:feature_post_processing}). We describe our particle-based formulation in the next section.

\section{Formulation}
\label{sec:method}

Given a set of seed spheres $\{\msphere_i=(\mcenter_i, r_i)\}_{i=1}^n$ as initialization described in Sec.~\ref{sec:opt_init}, our goal is to optimize their distribution in a structure-aware manner, ensuring that centers are evenly placed along each medial sub-structure (\ie sheet or seam) in the medial mesh $\mmesh$, while also aligning them with the medial axis through projection (Sec.~\ref{sec:tech_sph_proj}). We promote even distribution by minimizing a particle-based repulsion energy (Sec.~\ref{sec:method_particle}). However, we do not allow the repulsion to move spheres arbitrarily as we restrict each sphere’s movement to the sub-structure it belongs to. To achieve this, we constrain the optimization gradients using the solution space derived from RPD-based \textit{Spherical Quadratic Error Metrics} (SQEM) (Sec.~\ref{sec:method_qem}, and Sec.~\ref{sec:particle_grad_proj}). After each optimization step, spheres are projected back onto the medial axis to reinforce structural alignment.

It is worth mentioning that our method adopts a topology preservation strategy similar to MATTopo~\cite{wang2024mattopo} and requires the computation of a volumetric RPD. As such, we follow the same input assumption: the input must be a manifold tetrahedral mesh with a single connected component, no self-intersections, and no internal cavities. Notably, this `no cavity' assumption is used to preserve topological consistency in the RPD framework, and is satisfied by all CAD and organic models encountered in our experiments.

\begin{figure}[!h]
\centering
\begin{minipage}[t]{0.48\linewidth}
    \centering
    \includegraphics[width=\linewidth]{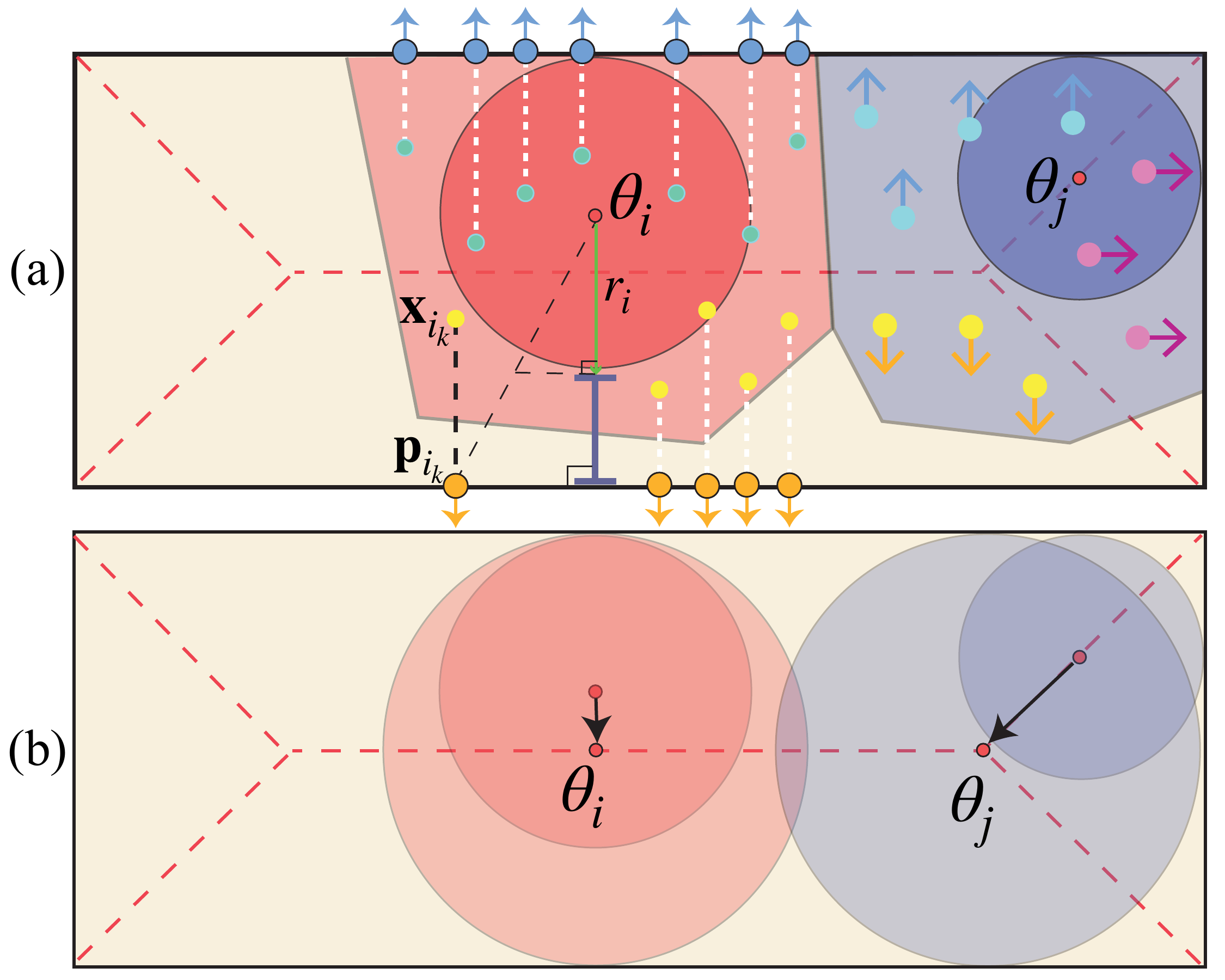}
    \caption{Illustration of our RPD-based spherical quadratic error metric described in Sec.~\ref{sec:method_qem}.}
    \label{fig:method_qem}
\end{minipage} 
\hfill
\begin{minipage}[t]{0.49\linewidth} 
    \centering
    \includegraphics[width=\linewidth]{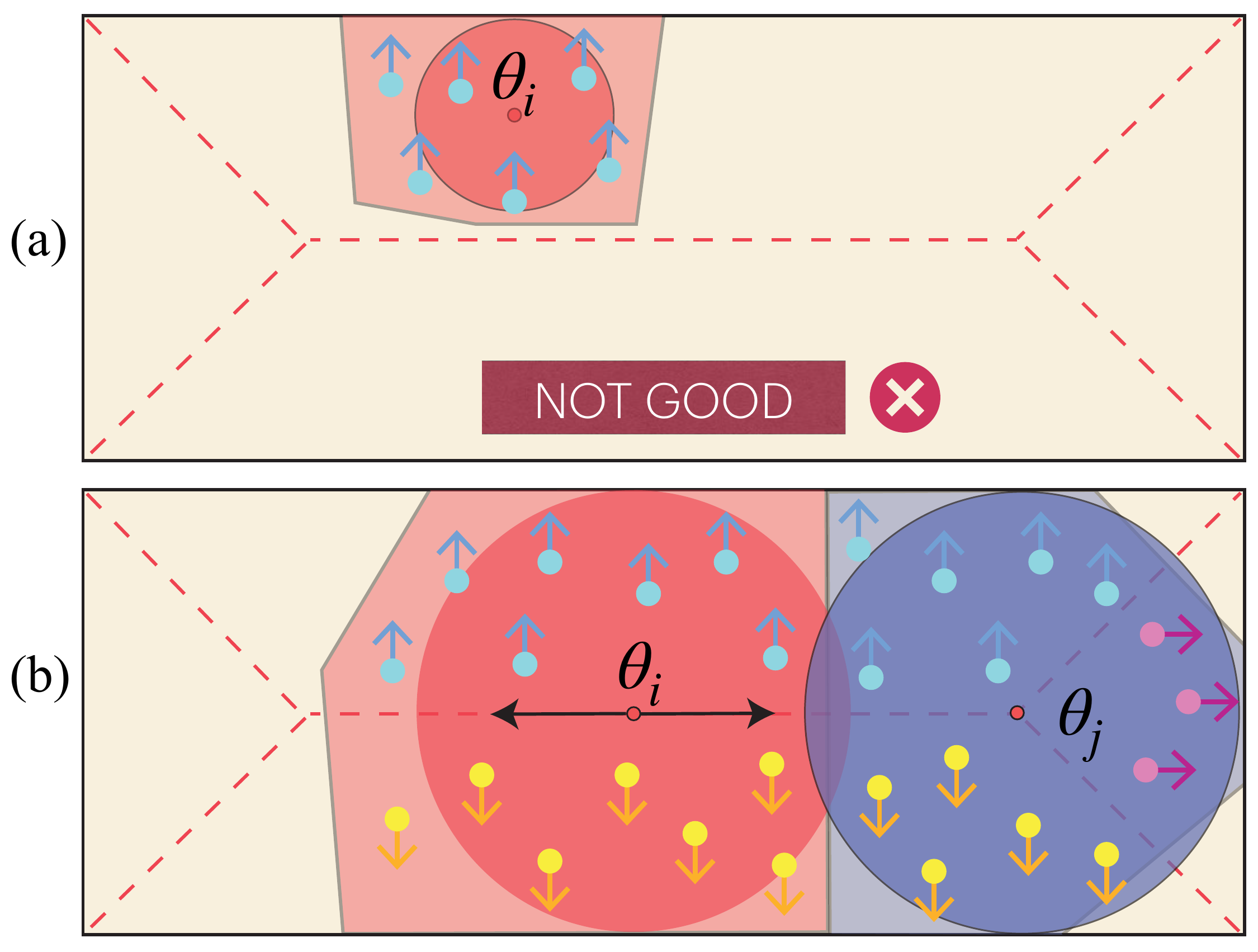}
    \caption{Illustration of our gradient projection strategy described in Sec.~\ref{sec:particle_grad_proj}.}
    \label{fig:energy_particle_proj}
\end{minipage}
\end{figure}

\subsection{Particle Function}
\label{sec:method_particle}

Crowding of spheres can lead to numerical instability when preserving medial structures (see Fig.~\ref{fig:intro_why}), while a more uniform distribution also improves triangle quality in the final mesh. To optimize sphere distribution, we adopt a particle-based repulsion function inspired by Witkin and Heckbert~\shortcite{witkin1994using} and Zhong et al.~\shortcite{zhong2013particle}.
Given $n$ spheres with centers $\{\mcenter_i\}_{i=1}^n$, which are treated as particles without radii, we define the \textit{inter-particle energy} between particles $i$ and $j$ as $\energy_{par}^{ij} = e^{-\frac{||\mcenter_j - \mcenter_i||^2}{2\kernel^2}}$,
where $\kernel$, referred to as the \textit{kernel width}, is the fixed standard deviation of the Gaussian kernel. We discuss the selection of an appropriate $\kernel$ in Sec.~\ref{sec:opt_particle}. Notably, $\energy_{par}^{ij} = \energy_{par}^{ji}$.  

When the forces acting on each particle reach equilibrium, the particles achieve an optimal balanced state with a uniform distribution. The gradient of $\energy_{par}^{ij}$ with respect to $\mcenter_i$ can be interpreted as the force $F_{par}^{ij}$ exerted on particle $i$ by particle $j$:
\begin{equation}
F_{par}^{ij} = \frac{\partial \energy_{par}^{ij}}{\partial \mcenter_j} = \frac{(\mcenter_j - \mcenter_i)}{\kernel^2} e^{-\frac{||\mcenter_j - \mcenter_i||^2}{2\kernel^2}}.
\label{eqn:particle_force}
\end{equation}
By minimizing the total energy 
\begin{equation}
\energy_{par} = \sum_i \sum_{j, j \neq i} \energy_{par}^{ij}
\label{eqn:particle}
\end{equation}
using the L-BFGS optimization algorithm \cite{zhong2013particle}, we can achieve a uniform sampling of sphere centers. 

However, allowing particles to move freely inside the shape without spatial constraints or structure awareness can lead to undesirable behavior, which will be elaborated more in Sec.~\ref{sec:particle_grad_proj}. To address this, we introduce an RPD-based formulation using the \textit{spherical quadric error metric} (SQEM)~\cite{thiery2013sphere}, which models the squared distance from a sphere to a set of planes.

\subsection{RPD-based Spherical Quadratic Error Metrics}
\label{sec:method_qem}
The RPD-based SQEM equation for a given sphere $\msphere_i$ is defined as:
\begin{equation}
    \text{SQEM}^i = \sum_{\Omega_i} [(\anypointproj_{i_k} - \mcenter_i)^{\top} \anypointnormal_{i_k} - r_i]^2,
\label{eqn:sqem}
\end{equation}
where $\anypointproj_{i_k}$ is the surface projection of the sample $\anypoint_{i_k}$ within the RPC $\rpc(\msphere_i)$ and  $\anypointnormal_{i_k} = \frac{\anypointproj_{i_k} - \anypoint_{i_k}}{||\anypointproj_{i_k} - \anypoint_{i_k}||}$ represents the direction from \(\anypoint_{i_k}\) to its projection $\anypointproj_{i_k}$. Minimizing this equation is equivalent to solving a linear least squares problem, which can be expressed as 
\begin{equation}
    \argmin_{\msphere_i}||A\msphere_i - b||^2 ,
\label{eqn:sqem_ls}
\end{equation}
where $\msphere_i$ is a 4D vector representing the sphere’s center and radius $(\mcenter_i, r_i)$, and $A$ is a $4 \times 4$ matrix.

Take a 2D rectangle as an example, see Fig.~\ref{fig:method_qem}, where the GT medial axis is shown in dotted red lines. 
For a sphere $\msphere_i$, shown in red, its RPC touches two sub-volumes within local region, upper and lower, divided by the GT medial axis. 
For each sample $\anypoint_{i_k}$ 
within an RPC cell, its closest point on the shape’s surface, $\anypointproj_{i_k}$, can be identified. This allows us to assign a normal vector $\anypointnormal_{i_k}$ to $\anypoint_{i_k}$, pointing from $\anypoint_{i_k}$ to $\anypointproj_{i_k}$. The RPD-based SQEM defined at vertex $\anypoint_{i_k}$ measures the distance from the sphere to the plane defined by $\anypointproj_{i_k}$ and the normal $\anypointnormal_{i_k}$. 
The total SQEM equation across all samples in an RPC cell quantifies how close the sphere is to the shape’s surface, using volumetric information. 
Its minimizer yields the sphere that best fits those planes in the least-squares sense.
Similarly, for a junction sphere $\msphere_j$ (shown in blue sphere in Fig.~\ref{fig:method_qem}), its RPC spans multiple local sub-volumes (upper, lower and right). By minimizing the SQEM equation, the sphere is repositioned to the position that ensures proper tangency with multiple surface planes.

Even though the solution space of the RPD-based SQEM minimizer does not exactly align with the definition of medial spheres (which are tangent to at least two surface points rather than to a set of planes), it provides a good approximation of the feasible space of medial spheres. We leverage this approximation to constrain particle movement, as described in the next section.

\subsection{Structure-Aware Gradient Projection}
\label{sec:particle_grad_proj}

As described above, we aim to prevent particles from moving freely inside the shape without spatial constraints or structure awareness.
For example, in Fig.~\ref{fig:energy_particle_proj}~(a), consider a particle $\theta_i$ that has moved far from the medial axis, such that the RPC of the corresponding sphere lies entirely within the upper sub-volume. To avoid such issues, we restrict particle movements using RPD spatial information by analyzing the singular values of the matrix $A$ described in Sec.~\ref{sec:method_qem}.
We denote the four singular values of $A$ as $\singular_1 \geq \singular_2 \geq \singular_3 \geq \singular_4 \geq 0$. The solution space of $\msphere_i$ depends on these singular values:

\vspace{-5pt}
\paragraph{Case 1}
If all singular values are significantly greater than $0$, there is no degree of freedom, and $\msphere_i$ can be directly solved from Eq.~\ref{eqn:sqem_ls}. 
A 2D example is shown in Fig.~\ref{fig:energy_particle_proj} (b), where the blue sphere with center $\mcenter_j$ has an RPC that touches three sub-volumes (upper, lower, and right), fully constraining all degrees of freedom. The projected gradient is set to zero, effectively fixing the sphere for that iteration.

\vspace{-5pt}
\paragraph{Case 2}
If only $\singular_4 \approx 0$, the solution space of $\msphere_i$ is constrained to a line space with one degree of freedom. Geometrically, this means the sphere can only move along a specific line. We compute the tangent vector $\mathbf{v}$ of this line and project the gradient $\mathbf{g_i} = \sum_{j=1}^k F_{par}^{ij}$ onto it as $\mathbf{g_i}^{proj} = \mathbf{v}\mathbf{v}^{\top} \mathbf{g_i}$. 
A 2D example is shown in Fig.~\ref{fig:energy_particle_proj} (b), where the red sphere with center $\mcenter_i$ has an RPC that intersects both the upper and lower sub-volumes, indicating that its movement is restricted to a specific line.

\vspace{-5pt}
\paragraph{Case 3}
If $\singular_{3,4} \approx 0$, the solution space is a plane. We derive this plane from the corresponding singular vectors and obtain its normal $\mathbf{n}$. Similar to Case 2, we project the gradient $\mathbf{g_i}$ onto the plane using $\mathbf{g_i}^{proj} = \mathbf{g_i} - \mathbf{n}\mathbf{n}^{\top} \mathbf{g_i} $.

\vspace{-5pt}
\paragraph{Case 4}
If $\singular_{2,3,4} \approx 0$, the seed sphere is too far from the medial axis, and its RPC does not sufficiently constrain its movement. In this case, we set the projected gradient to zero and use the \textit{sphere-shrinking} algorithm~\cite{ma20123shrink}, see Sec.~\ref{sec:tech_sph_proj}, to relocate the seed sphere onto the medial axis.

By projecting the gradient of the particle-based energy function onto the null space of its SQEM minimizer, we ensure that spheres move only in permissible directions while optimizing the even distribution within each sub-structure.
The null space of SQEM minimizer includes both center and radius dimensions. In our implementation, we constrain the gradient $g_i$ to the first three components (center) and discard the radius. The radius is updated separately (Sec.~\ref{sec:opt_particle}), which prevents it from becoming negative, though centers may still drift outside. To address this, we apply a correction step at the end of each iteration (Sec.~\ref{sec:tech_sph_proj}), projecting such spheres back onto the medial sheet. We also discuss the SQEM numeral stability in \supmaterial{Sec.~2} (Supplementary Material).

\subsection{Medial Sphere Projection}
\label{sec:tech_sph_proj}

After each particle optimization, the updated seed spheres ${\msphere_i}$ are projected to their nearest locations on the medial axis $\ma$. We avoid performing this projection during each particle iteration, as it introduces abrupt changes in the spheres' positions, potentially causing the particle energy to spike. This energy increase may exceed the energy prior to projection, thereby disrupting the L-BFGS optimization and hindering gradient descent direction finding.

We employ three projection strategies. First, we solve the SQEM in Eq.~\ref{eqn:sqem} as a linear system. For infinite solution spaces (Cases 2 and 3 in Sec.~\ref{sec:particle_grad_proj}), we project the current sphere onto the solution line (Case 2) or solution plane (Case 3) to obtain the closest feasible solution. While SQEM minimizes the squared distance from a sphere to a set of tangent planes in a linear least-squares sense, a medial sphere by definition must be tangent to at least two surface points. Hence, minimizing SQEM alone may not yield accurate medial spheres.
To address this, our second projection strategy applies the \textit{sphere-optimization} algorithm~\cite{2022MATFP}, which explicitly enforces point-tangent constraints. Please refer to \supmaterial{Sec.~3} (Supplementary Material) for details.

However, when a seed sphere deviates too far from the medial axis (Case 4), even sphere-optimization may fail, especially if the sphere drifts outside the shape. As a fallback, we employ the \textit{sphere-shrinking} algorithm~\cite{ma20123shrink} 
to project such spheres back onto the medial axis (see Fig.~\ref{fig:post_rpc_shrink} left). Given a pin point $\mathbf{p}$ with normal $\mathbf{n}_p$ on the boundary $\bmodel$, the algorithm iteratively reduces the sphere radius until the sphere $\msphere^t$ becomes a maximal empty ball, maintaining another tangent point $\mathbf{q}^t$ with normal $\mathbf{n}_q^t$.

\section{Optimization \& Technical Details}
\label{sec:opt}

In this section, we present the technical details of our optimization framework. We begin with the initialization strategy in Sec.~\ref{sec:opt_init}, followed by the particle-based optimization in Sec.~\ref{sec:opt_particle}, where we explain the design choice of Gaussian kernels. The complete optimization algorithm is outlined in Alg.~\ref{alg:opt}.

\begin{algorithm}
\caption{Structure-aware Particle Optimization}
\label{alg:opt}
\KwData{input tetrahedral mesh}
\KwResult{medial mesh $\mmesh$}
Initialize spheres $\{\msphere_i\}$ in Sec.~\ref{sec:opt_init}; \\
Run medial feature preservation in \supmaterial{Sec.~4} (Supplementary Material);\\
\While{medial structure not converged}{
    \While{stopping condition not satisfied}{
        compute restricted power diagram (RPD);\\
        Sample each restricted power cell (RPC) of $\{\msphere_i\}$;  \\
        \For{each sphere center $\msphere_i$} {
            \For{each neighbors $\msphere_j$ of $\msphere_i$}{
                Compute $\energy_{par}^{ij}$; \\
                Compute $F_{par}^{ij}$ using Eq.~\ref{eqn:particle_force};
            }
            Sum the total force $F_{par}^{i}$; \\
            Project $F_{par}^{i}$ in Sec.~\ref{sec:particle_grad_proj};
        }
        Sum the total energy $\energy_{par}$ using Eq.~\ref{eqn:particle}; \\
        Run L-BFGS with $\energy_{par}$ and $\{F_{par}^{i}\}$; \\
    }
    Project spheres $\{\msphere_i\}$ in Sec.~\ref{sec:tech_sph_proj}; \\
    Run medial feature preservation in \supmaterial{Sec.~4} (Supplementary Material);\\
}
Compute medial mesh as the dual of \text{RPD}; \\
Run post-process operation in \supmaterial{Sec.~5} (Supplementary Material);
\end{algorithm}

\vspace{-5pt}
\subsection{Initialization}
\label{sec:opt_init}
Since our optimization restricts particle movement across different sheets, the initial medial spheres are uniformly sampled based on the shape surface $\bmodel$ to maximize the coverage of each individual sheet structure. To sample $T_2$ spheres on the sheets of the medial structure, we employ the \textit{sphere-shrinking} algorithm~\cite{ma20123shrink}, which generates spheres tangent to two distinct surface regions. 
Specifically, we use Poisson-disk sampling to distribute `pin' points on $\bmodel$, allowing the user to adjust the sampling parameter $\gamma$. The sampling radius is defined as $\frac{1}{\gamma}$ times the model's bounding box diagonal length. We use $\gamma=40$ in our experiment and provide an ablation study in \supmaterial{Sec.~7} (Supplementary Material).

\subsection{Optimizing Particle Function}
\label{sec:opt_particle}
For the inter-particle energy defined in Eq.~\ref{eqn:particle}, we adopt a fixed kernel width $\kernel$, following Zhong et al.~\shortcite{zhong2013particle}. In our approach, $\kernel$ is proportional to the average `radius' of particles when uniformly distributed on the initial medial mesh $\mmesh^0$: $\kernel = c_{\kernel} \sqrt{|\mmesh^0| / n}$, where $|\mmesh^0|$ represents the area of the initial medial spheres (Sec.~\ref{sec:opt_init}), $n$ is the number of initial medial spheres, and $c_{\kernel}$ is a constant coefficient set to $0.3$ in all experiments. We use the L-BFGS algorithm~\cite{liu1989limited} to optimize particle positions. In each L-BFGS iteration, we update the total energy $\energy_{par}$ (Eq.~\ref{eqn:particle}) and compute the total force acting on each particle $F_{par}^i = \sum_{j=1}^{\text{Neigh}} F_{par}^{ij}$ using KNN where $\text{Neigh}=10$ in our experiments. 
Additionally, at each iteration, we update the sphere radius to be the nearest distance to the model surface $\bmodel$, as only sphere centers are adjusted during optimization. Here we show the pseudocode for our optimization strategy in Alg.~\ref{alg:opt}. 
We terminate the outer loop in Alg.~\ref{alg:opt} when the medial structure has converged. Specifically, we monitor changes in the number of seam and junction spheres, and stop the loop when the changes' ratio falls below a predefined threshold (set to $3 \times 10^{-4}$ in our experiments) or after a maximum number of iterations (set to $30$ in our experiments) The inner loop is terminated when the magnitude of the gradient drops below a specified threshold (set to $5 \times 10^{-3}$ in our experiment).

\begin{figure}[h]
	\centering
	\includegraphics[width=\linewidth]{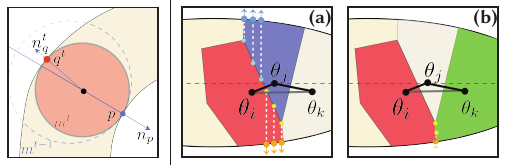}
    \vspace{-10pt}
    \caption{Left: the \textit{sphere-shrinking} algorithm~\cite{ma20123shrink} used in Sec.~\ref{sec:tech_sph_proj}. Right: Illustration of pruning invalid medial connections described in Sec.~\ref{sec:feature_post_processing}.}
	\label{fig:post_rpc_shrink}
    \vspace{-10pt}
\end{figure}

\subsection{Medial Feature Preservation and Post-Processing}
\label{sec:feature_post_processing}

As our method leverages the restricted power diagram (RPD) to decompose the input volume, similar to MATFP~\cite{2022MATFP} and MATTopo~\cite{wang2024mattopo}, we adopt comparable strategies for medial feature preservation and post-processing including \textit{topology preservation} and \textit{structure-aware thinning process}. Additional implementation details are provided in \supmaterial{Sec.~4} and \supmaterial{Sec.~5} Supplementary Material.

\paragraph{Pruning Invalid Medial Connections}
To ensure a clean medial mesh structure, we prioritize pruning medial edges and faces that span across different sheets during the thinning process. For each medial edge $\medge_{ij}$ connecting spheres $\msphere_i$ and $\msphere_j$, we examine its dual restricted power face (RPF). If the RPF intersects the same sub-volumes as at least one of its endpoint RPCs, the edge is considered valid; otherwise, it is pruned. Fig.~\ref{fig:post_rpc_shrink} right (a) shows a valid 2D example where samples on its dual structure spans both upper and lower sub-volumes. In contrast, Fig.~\ref{fig:post_rpc_shrink} right (b) shows an invalid edge whose dual samples lie entirely within one sub-volume. A medial face $\mface_{ijk}$ is considered invalid if any of its bounding edges are invalid.

\section{Experiments}
\label{sec:exp}
We present both quantitative and qualitative evaluations of the proposed method.
All experiments are conducted on a machine equipped with a 3.60GHz Intel(R) Core(TM) i7-9700K CPU and 32 GB of RAM.
We evaluate our method on a total of $100$ CAD models from the ABC dataset~\cite{koch2019abc, xu2024cwf, xu2022rfeps}, as well as $14$ organic models exhibiting diverse topologies. All input models are normalized to fit within the $[0, 1000]^3$ bounding box. For tetrahedral meshing, we use fTetWild~\cite{hu2020ftetwild} with a target edge length parameter $l = 0.5$. We conduct an ablation study in \supmaterial{Sec.~7} (Supplementary Material) to evaluate how different values of $l$, which control the tessellation quality, affect our results. The $\#s$ is the number of generated medial spheres.

\paragraph{Comparison Methods}
We compare our method against six state-of-the-art (SOTA) approaches: PC~\cite{amenta2001power}, SAT~\cite{miklos2010sat}, VC~\cite{yan2018voxel}, MATFP~\cite{2022MATFP}, MATTopo~\cite{wang2024mattopo}, and VMAS~\cite{vmas2024}. Among these, VMAS is a skeletonization method designed to generate sparse medial representations (typically a few hundred spheres), while the others are medial axis approximation methods.
We include VMAS in our comparison due to its variational formulation. However, VMAS becomes unstable when the target number of spheres increases to several thousand. In such cases, the optimization exhibits oscillatory behavior as spheres are repeatedly inserted and deleted without convergence.
For a fair comparison, we run VMAS on dense surface meshes (each with over $300\mathrm{k}$ vertices) and incrementally increase the target number of spheres. We terminate the process if the number of spheres does not change for $100$ consecutive iterations, which allows us to evaluate VMAS under its maximum stable capacity.

\paragraph{Evaluation Metrics}
To evaluate the quality of the generated medial mesh, we employ three indicators: \textit{Medial Structure Error Ratio} (MSER), \textit{Triangle Quality} (\triangleQ)~\cite{frey1999surface}, and \textit{Topology Error Ratio} (\textit{TER}) (incorrect Euler characteristic ratio). To assess the geometric fidelity between the reconstructed surface and the original input surface, we use the \textit{Hausdorff Distance} ($\hd$) as the evaluation metric. Due to page limits, detailed descriptions of \triangleQ, TER, and $\hd$ are provided in \supmaterial{Sec.~6} (Supplementary Material).

\paragraph{Medial Structure Error Ratio (MSER)}
\label{sec:exp_eval_mser}

\begin{figure}[h]
    \centering
    \includegraphics[width=0.9\linewidth]{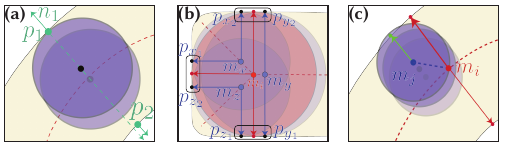}
    \vspace{-5pt}
    \caption{Illustration of medial structure error ratio (MSER) described in Sec.~\ref{sec:exp_eval_mser}.}
    \label{fig:exp_mser}
\end{figure}

To evaluate the accuracy of the computed medial structure, we introduce the \textit{Medial Structure Error Ratio} (MSER), which measures the percentage of misclassified seam and junction spheres. The key idea is that a sphere on a medial sheet should have exactly two distinct tangents on the shape surface. At seams—where multiple sheets intersect—the number of tangents should increase abruptly. MSER is designed to detect such structural transitions.
To compute MSER, we first recover tangents for each sheet sphere using a robust search strategy, as these spheres may not precisely at GT positions.
Starting from the closest surface point and normal, Fig.~\ref{fig:exp_mser} (a), we iteratively shift the sphere along the negative normal direction until a second valid tangent $\tanpoint_2$ is found or a maximum of 10 steps is reached, with step sized at $0.01$ times the sphere’s radius. 
Spheres that fail to find a second tangent (\eg $\msphere_j$ in Fig.~\ref{fig:exp_mser}~(c)) are marked as $T_1$ spikes, often arising from redundant sheets.
For each seam sphere, we aggregate tangents from neighboring sheet spheres, cluster them, and identify the closest surface point within each cluster region as its representative tangent, see $\msphere_i$ in Fig.~\ref{fig:exp_mser}~(b).  
If a seam sphere has three tangents (exceeding the tangent count of all its neighboring sheet spheres), it is correctly classified as a seam sphere; otherwise, it is considered misclassified. A misclassified example is shown in Fig.~\ref{fig:exp_mser} (c), where $\msphere_j$ lies on a redundant sheet (blue dotted line), and the seam sphere $\msphere_i$ (detected via non-manifoldness) has only two tangents and is therefore misclassified.
Junction spheres, where multiple seams intersect, are validated similarly by aggregating tangents from adjacent sheet spheres.
MSER is finally computed as the ratio of misclassified seam and junction spheres over the total number of such spheres.
We evaluate MSER on CAD models only, as they typically exhibit well-defined, clean geometric features characteristic of human-made designs, unlike the ambiguous structures often found in organic shapes.

\begin{table}[t!]
\caption{Quantitative comparison on $100$ CAD models, taken from the ABC dataset~\cite{koch2019abc}. The \underline{\textbf{best}} scores are emphasized in bold with underlining, while the \textbf{second best} scores are highlighted only in bold.}
\vspace{-10pt}
\begin{center}
\scalebox{0.7}{
\begin{tabular}{c|c||c|c|c|c|c|c|c}
\toprule
 Metrics &   & PC & SAT & VC & MATFP & MATTopo & VMAS & Ours \\ 
 \midrule
 \multirow{3}{*}{MSER~$\downarrow$} 
    & Avg   &- &- &\textbf{0.334} &0.437 &0.485 &- &\underline{\textbf{0.009}}  \\
    & 85th  &- &- &0.753 &\textbf{0.736} &0.864 &- &\underline{\textbf{0.008}}  \\
    & 90th  &- &- &0.826 &\textbf{0.808} &0.899 &- &\underline{\textbf{0.015}}  \\ 
    \midrule
\multirow{3}{*}{\triangleQ~$\uparrow$} 
    & Avg &0.421  &0.249  &0.599 &0.423 &0.485 &\textbf{0.699} &\underline{\textbf{0.712}} \\
    & 85th &0.473  &0.271  &0.667 &0.477 &0.523 &\textbf{0.762} &\underline{\textbf{0.766}} \\
    & 90th &0.483  &0.277  &0.681 &0.489 &0.535 &\textbf{0.773} &\underline{\textbf{0.777}} \\
    \midrule
TER~$\downarrow$ 
    &- &1.00 & 1.00 &0.03 &0.27 &\underline{\textbf{0.00}} &0.87 &\underline{\textbf{0.00}} \\ 
    \midrule
\midrule
\multirow{3}{*}{HD (\%)~$\downarrow$} 
    & Avg   &2.284 &1.223 &0.977 &\textbf{0.744} &0.805 &1.654  &\underline{\textbf{0.735}}  \\
    & 85th  &3.533 &1.354 &1.238 &\underline{\textbf{1.029}} &\textbf{1.058} &2.538  &1.194  \\
    & 90th  &4.272 &1.403 &1.432 &\underline{\textbf{1.114}} &\textbf{1.137} &3.161  &1.625  \\ 
\bottomrule
\end{tabular}}
\end{center}
\label{tab:exp_tab_cad}
\end{table}

\begin{table}[t!]
\small
\caption{Quantitative comparison on $14$ organic models. 
}
\vspace{-10pt}
\begin{center}
\scalebox{0.7}{
\begin{tabular}{c|c||c|c|c|c|c|c|c}
\toprule
 Metrics &   & PC & SAT & VC & MATFP & MATTopo & VMAS & Ours \\ \midrule
\multirow{3}{*}{\triangleQ~$\uparrow$} 
    & Avg & 0.391 & 0.224 & 0.415 & 0.431 & 0.496 & \textbf{0.594} &\underline{\textbf{0.691}}\\
    & 85th & 0.412 & 0.248 & 0.441 & 0.473 & 0.516 & \textbf{0.656} &\underline{\textbf{0.788}}\\
    & 90th & 0.413 & 0.252 & 0.469 & 0.475 & 0.519 & \textbf{0.667} &\underline{\textbf{0.808}}\\
    \midrule
TER~$\downarrow$ 
    &- & 1.00 & 1.00 & 0.07 & 1.00 & \underline{\textbf{0.00}} & 1.00 & \underline{\textbf{0.00}} \\ 
    \midrule
\midrule
\multirow{3}{*}{HD$(\%)$~$\downarrow$} 
    & Avg   & 2.355 & 1.011 & 0.971 & \underline{\textbf{0.608}} & 0.775 & 1.822 & \textbf{0.624}\\
    & 85th  & 2.728 & 1.168 & 1.289 & \textbf{1.047} & \underline{\textbf{1.027}} & 3.871 & 1.243\\
    & 90th  & 2.919 & 1.269 & 1.369 & \underline{\textbf{1.081}} & \textbf{1.108} & 4.578 & 1.348\\ 
\bottomrule
\end{tabular}}
\vspace{-5pt}
\end{center}
\label{tab:exp_tab_organic}
\end{table}

\subsection{Comparisons on CAD Models}

\begin{figure*}[!t]
    \centering
    \includegraphics[width=\linewidth]{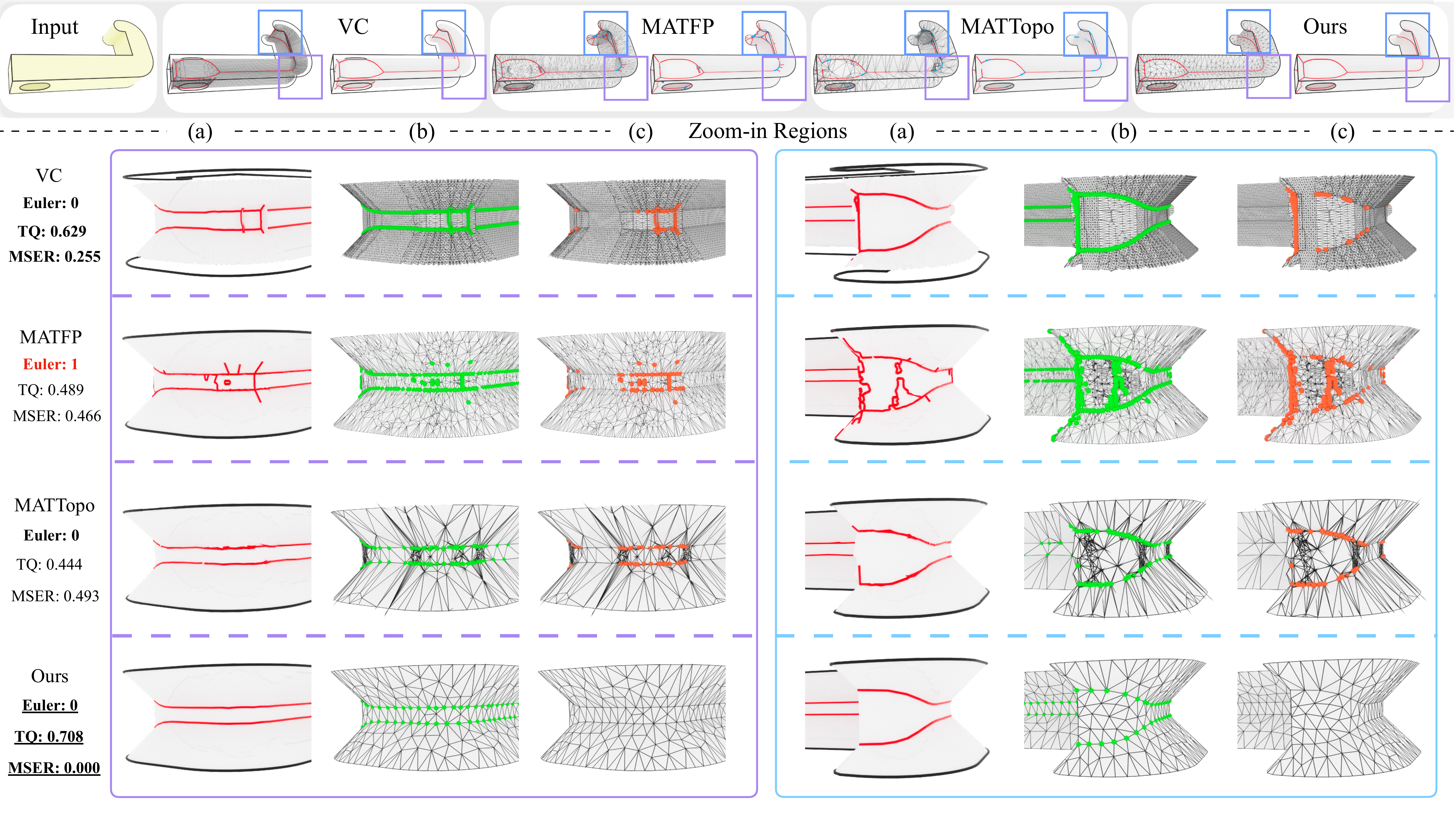}
    \vspace{-35pt}
    \caption{Comparison of the structure awareness of the generated medial mesh among VC~\cite{yan2018voxel}, MATFP~\cite{2022MATFP}, MATTopo~\cite{wang2024mattopo}, and our method. Detected medial seams are shown in red and junctions in blue. The zoom-in regions highlight: (a) medial seams; (b) medial spheres in green located on seams and junctions; and (c) misclassified spheres in orange according to the MSER metric (see Sec.~\ref{sec:exp_eval_mser}). Our method demonstrates improved structural awareness and higher triangle quality compared to state-of-the-art approaches.} 
    \label{fig:comp_mser_abc}
\end{figure*}

\begin{figure*}[h]
    \centering
    \includegraphics[width=0.9\linewidth]{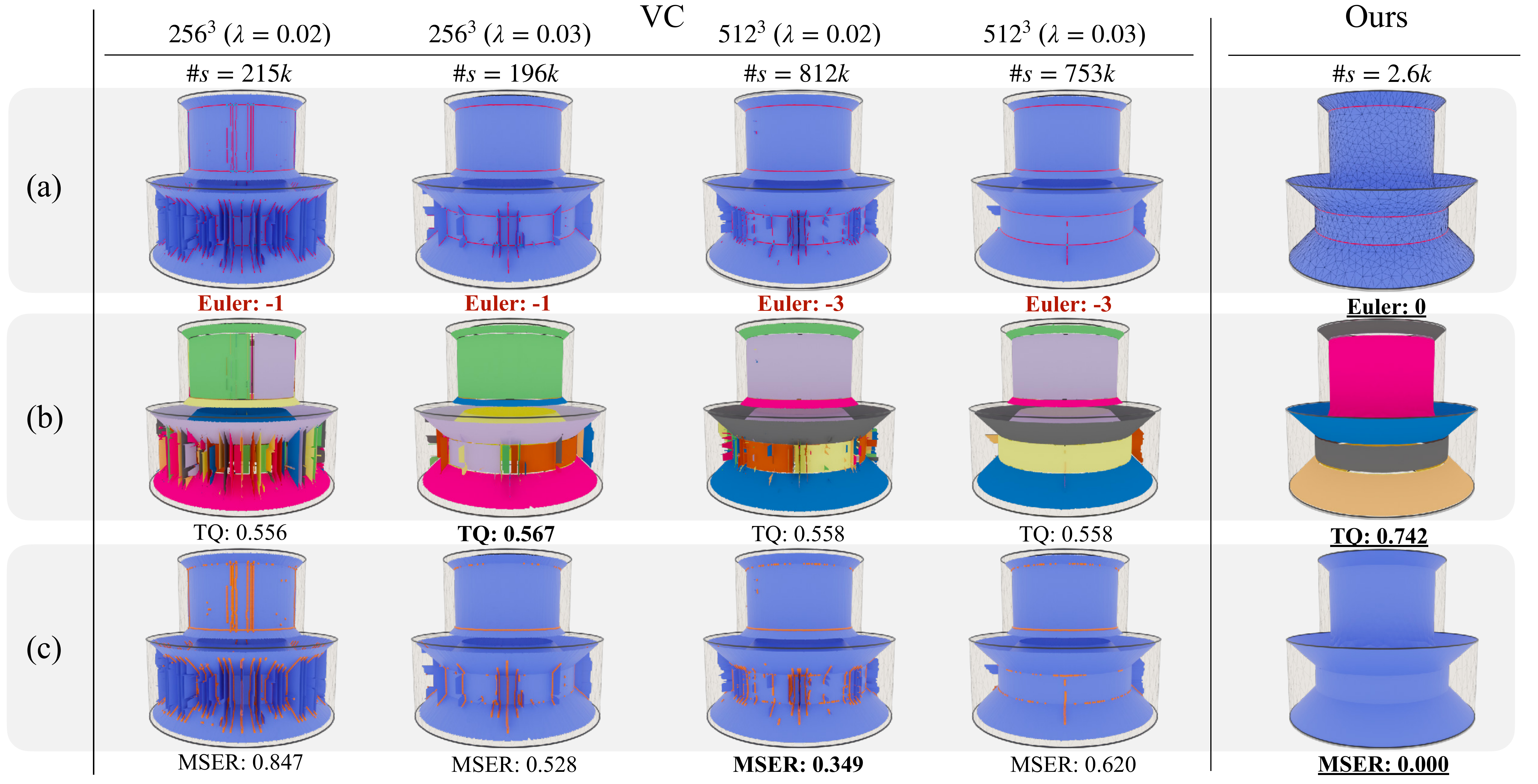}
    \vspace{-5pt}
    \caption{Comparison with VC~\cite{yan2018voxel} on medial structure using two different voxel resolutions ($2^8$ and $2^9$) and pruning parameters ($\lambda = 0.02$ and $\lambda = 0.03$). (a) Medial meshes with seams shown in red and junctions in blue. For VC, edges are omitted due to excessive density (over 500k edges). (b) Extracted medial sheets are visualized with different colors for each sheet. (c) Misclassified medial spheres are shown in orange. Our method produces cleaner medial structures with significantly higher triangle quality.}
    \label{fig:comp_vc}
\end{figure*}

We present the statistics of quantitative comparisons for the $100$ CAD models in Tab.~\ref{tab:exp_tab_cad}. 
Since PC~\cite{amenta2001power}, SAT~\cite{miklos2010sat}, and VMAS~\cite{vmas2024} produce medial meshes with 3-dimensional cells (`closed pockets'), violating the thinness property, we exclude them from medial structure evaluation.  Instead, we compare our method with VC~\cite{yan2018voxel}, MATFP~\cite{2022MATFP}, and MATTopo~\cite{wang2024mattopo} using the MSER metric. For VC, seams and junctions are extracted via non-manifold analysis, where seams are defined as intersections of multiple manifold sheets, and junctions as intersections of multiple seams. We use a voxel resolution of $2^8$ and a pruning parameter of $\lambda=0.03$. For MATFP and MATTopo, we adopt their original sphere classification and feature extraction strategies. For medial meshes with relatively few spheres, such as those generated by MATTopo and ours, which typically contain a few thousand spheres (as opposed to hundreds of thousands in VC), we perform mesh subdivision prior to evaluation to ensure sufficient sampling on medial sheets for a fair comparison. Note that the newly added samples during subdivision are excluded from the MSER computation.
Qualitative and quantitative comparisons using MSER are shown in Fig.~\ref{fig:comp_mser_abc}. An additional comparison with VC under a different pruning setting ($\lambda=0.02$) is shown in Fig.~\ref{fig:comp_vc} to assess structure awareness. Reconstruction quality is evaluated in Fig.~\ref{fig:comp_tq_hd_abc_column}. Our method achieves better structural awareness and triangle quality while maintaining comparable reconstructability to MATFP and MATTopo, with the added benefit of preserving external features and ensuring topological consistency.

\subsection{Comparison on Organic Models}
\begin{figure*}[!h]
    \centering
    \begin{minipage}[t]{0.48\linewidth}
        \centering
        \includegraphics[width=\linewidth]{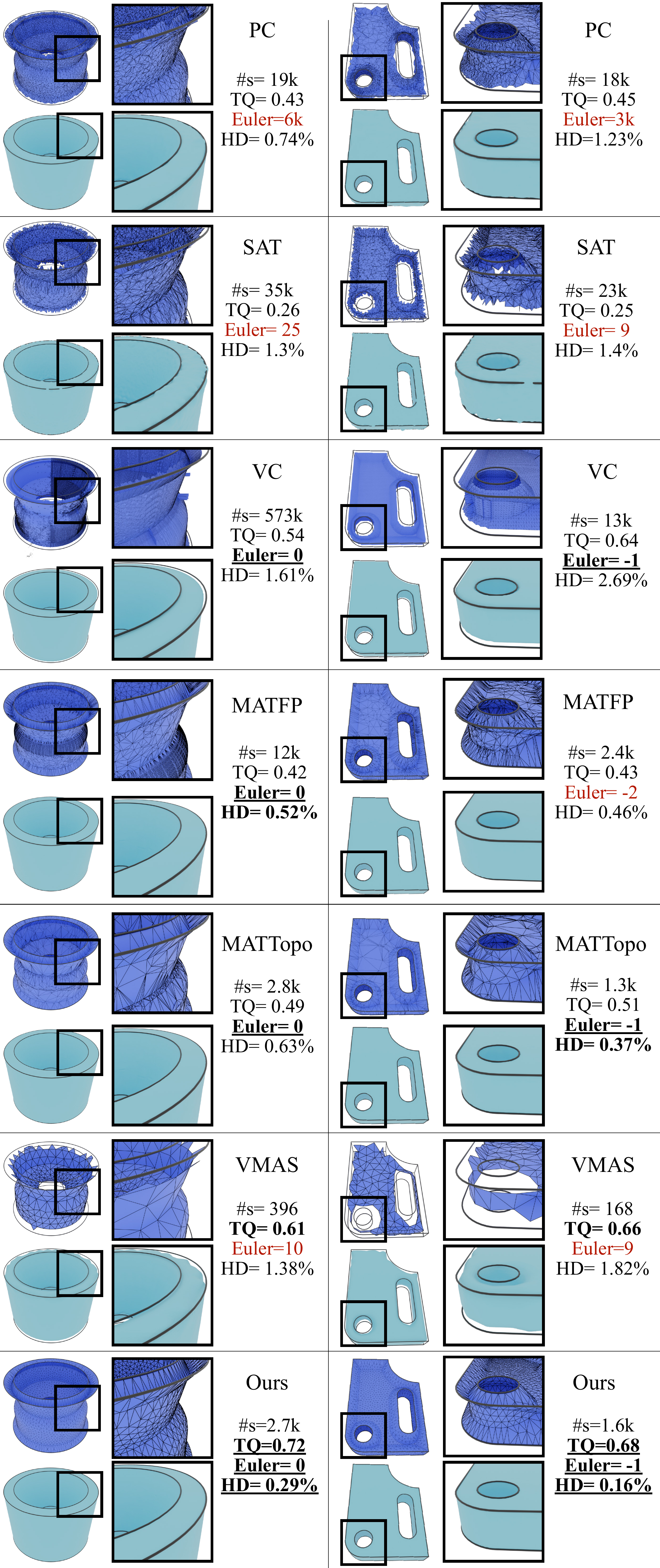}
    \vspace{-20pt}
    \caption{Comparison of the medial mesh and reconstructed surface with state-of-the-art methods on two CAD models. Our method achieves comparable reconstructability to MATFP~\cite{2022MATFP} and MATTopo~\cite{wang2024mattopo}, while offering improved structure awareness and higher triangle mesh quality.}
    \label{fig:comp_tq_hd_abc_column}
    \end{minipage} 
    \vspace{-30pt}
    \hfill
    \begin{minipage}[t]{0.48\linewidth} 
        \centering
        \includegraphics[width=\linewidth]{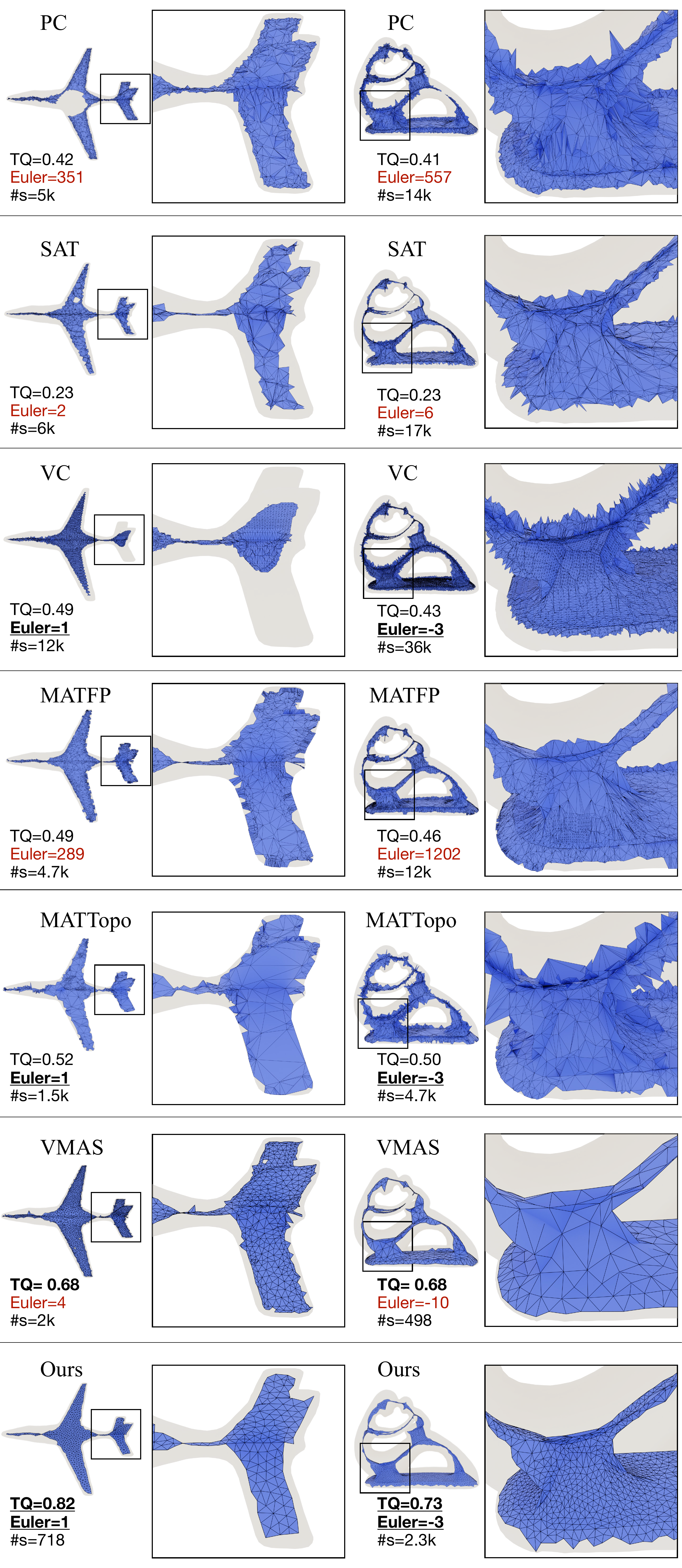}
    \vspace{-20pt}
    \caption{Comparison of medial meshes generated by state-of-the-art methods on two organic shapes. Our method more effectively consolidates structural features and produces high-quality medial meshes with improved connectivity, cleaner structures, and fewer artifacts.}
    \label{fig:comp_organic_column}
    \end{minipage}
\end{figure*}

We further compare our method with six state-of-the-art methods on $14$ organic shapes and report the quantitative results in Tab.~\ref{tab:exp_tab_organic}. Visual comparisons are provided in Fig.~\ref{fig:comp_organic_column}. Our method consistently achieves the best performance in triangulation quality and topology preservation, while attaining near-optimal reconstruction accuracy across all models.

\begin{table}[!h]
\small
\caption{
Statistics of our running time in seconds. $\#t$ denotes the number of tetrahedra in the input mesh, $\#s$ is the number of generated medial spheres, and $\#\text{itr}$ is the total number of optimization iterations. $S_{\text{Sample}}$ represents the total time spent sampling and projecting RPCs for all medial spheres across all iterations, while $S_{\text{RPD}}$ is the total time for computing the volumetric RPD.
}
\vspace{-10 pt}
\begin{center}
\scalebox{0.8}{
\begin{tabular}{c||c|c|c|c|c|c}
\hline
Model & $\#t$ & $\#s$ & $S_{\text{Sample}}$ (s) & $S_{\text{RPD}}$ (s) & Total (s) & Figure Reference
\\
\hline
7081 & 1.2k & 846 & 66 & 12 & 360 & Fig.~\ref{fig:comp_mser_abc}\\
\hline
40578 & 4.7k & 2.6k & 323 & 54 & 480 & Fig.~\ref{fig:comp_vc}\\
\hline
40049 & 5.6k & 2.7k & 337 & 55 & 540 & Fig.~\ref{fig:comp_tq_hd_abc_column}\\
\hline
40950 & 2.2k & 1.6k & 337 & 55 & 240 & Fig.~\ref{fig:comp_tq_hd_abc_column}\\
\hline
plane & 9.9k & 718 & 177 & 121 & 328 & Fig.~\ref{fig:comp_organic_column}\\
\hline
fertility & 8.8k & 2.3k & 336 & 120 & 600 & Fig.~\ref{fig:comp_organic_column}\\
\hline
metatron & 9k & 5.7k & 853 & 130 & 1020 & Fig.~\ref{fig:teaser} \\
\hline
\end{tabular}}
\end{center}
\label{tab:runtime}
\end{table}

\vspace{-10pt}
\section{Conclusion and Future Work}
\label{sec:limitations}
In conclusion, we present a novel framework for computing  structure-aware medial axis transform using particle optimization constrained by RPD-based SQEM. 
More analysis, ablation studies and limitations are provided in the \supmaterial{Supplementary Material}.
Our method focuses on generating structure-aware medial meshes with high triangle quality, while preserving external features and ensuring topological equivalence. However, this comes at the cost of increased computational complexity. In addition to computing the volumetric RPD, our method samples RPCs for each medial sphere and projects those samples onto the input surface. As a result, the overall runtime is higher compared to prior RPD-based methods such as MATFP~\cite{2022MATFP} and MATTopo~\cite{wang2024mattopo}.
As shown in Tab.~\ref{tab:runtime}, our method typically requires several minutes per model. The runtime is primarily influenced by the number of tetrahedra in the input mesh ($\#t$) and the number of medial spheres ($\#s$). In future work, we aim to parallelize the RPC sampling step ($S_{\text{Sample}}$) to significantly reduce computational time.
Moreover, similar to other variational partitioning algorithms, our method does not provide theoretical guarantees on global convergence or optimality. 

\begin{acks}
We thank all the anonymous reviewers for their insightful comments. Thanks also go to Shibo Song for helpful discussions. Zichun Zhong was partially supported by the National Science Foundation (OAC-1845962, OAC-1910469, and OAC-2311245). Rui Xu and Taku Komura were partially funded by the Research Grants Council of Hong Kong (Ref: 17210222).
\end{acks}

\newpage

\clearpage

\clearpage

\bibliographystyle{ACM-Reference-Format}
\bibliography{reference}


\begin{thebibliography}{58}


\ifx \showCODEN    \undefined \def \showCODEN     #1{\unskip}     \fi
\ifx \showDOI      \undefined \def \showDOI       #1{#1}\fi
\ifx \showISBNx    \undefined \def \showISBNx     #1{\unskip}     \fi
\ifx \showISBNxiii \undefined \def \showISBNxiii  #1{\unskip}     \fi
\ifx \showISSN     \undefined \def \showISSN      #1{\unskip}     \fi
\ifx \showLCCN     \undefined \def \showLCCN      #1{\unskip}     \fi
\ifx \shownote     \undefined \def \shownote      #1{#1}          \fi
\ifx \showarticletitle \undefined \def \showarticletitle #1{#1}   \fi
\ifx \showURL      \undefined \def \showURL       {\relax}        \fi
\providecommand\bibfield[2]{#2}
\providecommand\bibinfo[2]{#2}
\providecommand\natexlab[1]{#1}
\providecommand\showeprint[2][]{arXiv:#2}

\bibitem[Abdelkader et~al\mbox{.}(2020)]%
        {abdelkader2020vorocrust}
\bibfield{author}{\bibinfo{person}{Ahmed Abdelkader}, \bibinfo{person}{Chandrajit~L Bajaj}, \bibinfo{person}{Mohamed~S Ebeida}, \bibinfo{person}{Ahmed~H Mahmoud}, \bibinfo{person}{Scott~A Mitchell}, \bibinfo{person}{John~D Owens}, {and} \bibinfo{person}{Ahmad~A Rushdi}.} \bibinfo{year}{2020}\natexlab{}.
\newblock \showarticletitle{VoroCrust: Voronoi meshing without clipping}.
\newblock \bibinfo{journal}{\emph{ACM Transactions on Graphics (TOG)}} \bibinfo{volume}{39}, \bibinfo{number}{3} (\bibinfo{year}{2020}), \bibinfo{pages}{1--16}.
\newblock


\bibitem[Amenta et~al\mbox{.}(2001)]%
        {amenta2001power}
\bibfield{author}{\bibinfo{person}{Nina Amenta}, \bibinfo{person}{Sunghee Choi}, {and} \bibinfo{person}{Ravi~Krishna Kolluri}.} \bibinfo{year}{2001}\natexlab{}.
\newblock \showarticletitle{The power crust}. In \bibinfo{booktitle}{\emph{Proceedings of the sixth ACM symposium on Solid modeling and applications}}. \bibinfo{pages}{249--266}.
\newblock


\bibitem[Blum et~al\mbox{.}(1967)]%
        {blum1967transformation}
\bibfield{author}{\bibinfo{person}{Harry Blum} {et~al\mbox{.}}} \bibinfo{year}{1967}\natexlab{}.
\newblock \bibinfo{booktitle}{\emph{A transformation for extracting new descriptors of shape}}. Vol.~\bibinfo{volume}{43}.
\newblock \bibinfo{publisher}{MIT press Cambridge, MA}.
\newblock


\bibitem[Brandt and Algazi(1992)]%
        {brandt1992continuous}
\bibfield{author}{\bibinfo{person}{Jonathan~W Brandt} {and} \bibinfo{person}{V~Ralph Algazi}.} \bibinfo{year}{1992}\natexlab{}.
\newblock \showarticletitle{Continuous skeleton computation by Voronoi diagram}.
\newblock \bibinfo{journal}{\emph{CVGIP: Image understanding}} \bibinfo{volume}{55}, \bibinfo{number}{3} (\bibinfo{year}{1992}), \bibinfo{pages}{329--338}.
\newblock


\bibitem[Chang(2008)]%
        {chang2008medial}
\bibfield{author}{\bibinfo{person}{Ming-Ching Chang}.} \bibinfo{year}{2008}\natexlab{}.
\newblock \emph{\bibinfo{title}{The Medial Scaffold for 3D Shape Modeling and Recognition}}.
\newblock \bibinfo{thesistype}{Ph.\,D. Dissertation}. \bibinfo{school}{Ph. D. dissertation, Division of Engineering, Brown University, Providence~…}.
\newblock


\bibitem[Chang and Kimia(2008)]%
        {chang2008regularizing}
\bibfield{author}{\bibinfo{person}{Ming-Ching Chang} {and} \bibinfo{person}{Benjamin~B Kimia}.} \bibinfo{year}{2008}\natexlab{}.
\newblock \showarticletitle{Regularizing 3D medial axis using medial scaffold transforms}. In \bibinfo{booktitle}{\emph{2008 IEEE Conference on Computer Vision and Pattern Recognition}}. IEEE, \bibinfo{pages}{1--8}.
\newblock


\bibitem[Chazal and Lieutier(2005)]%
        {chazal2005lambda}
\bibfield{author}{\bibinfo{person}{Fr{\'e}d{\'e}ric Chazal} {and} \bibinfo{person}{Andr{\'e} Lieutier}.} \bibinfo{year}{2005}\natexlab{}.
\newblock \showarticletitle{The “$\lambda$-medial axis”}.
\newblock \bibinfo{journal}{\emph{Graphical Models}} \bibinfo{volume}{67}, \bibinfo{number}{4} (\bibinfo{year}{2005}), \bibinfo{pages}{304--331}.
\newblock


\bibitem[Chazal and Lieutier(2008)]%
        {chazal2008smooth}
\bibfield{author}{\bibinfo{person}{Fr{\'e}d{\'e}ric Chazal} {and} \bibinfo{person}{Andr{\'e} Lieutier}.} \bibinfo{year}{2008}\natexlab{}.
\newblock \showarticletitle{Smooth manifold reconstruction from noisy and non-uniform approximation with guarantees}.
\newblock \bibinfo{journal}{\emph{Computational Geometry}} \bibinfo{volume}{40}, \bibinfo{number}{2} (\bibinfo{year}{2008}), \bibinfo{pages}{156--170}.
\newblock


\bibitem[Chen and Holst(2011)]%
        {chen2011efficient}
\bibfield{author}{\bibinfo{person}{Long Chen} {and} \bibinfo{person}{Michael Holst}.} \bibinfo{year}{2011}\natexlab{}.
\newblock \showarticletitle{Efficient mesh optimization schemes based on optimal Delaunay triangulations}.
\newblock \bibinfo{journal}{\emph{Computer Methods in Applied Mechanics and Engineering}} \bibinfo{volume}{200}, \bibinfo{number}{9-12} (\bibinfo{year}{2011}), \bibinfo{pages}{967--984}.
\newblock


\bibitem[Cl{\'e}mot and Digne(2023)]%
        {clemot2023neural}
\bibfield{author}{\bibinfo{person}{Matt{\'e}o Cl{\'e}mot} {and} \bibinfo{person}{Julie Digne}.} \bibinfo{year}{2023}\natexlab{}.
\newblock \showarticletitle{Neural skeleton: Implicit neural representation away from the surface}.
\newblock \bibinfo{journal}{\emph{Computers \& Graphics}}  \bibinfo{volume}{114} (\bibinfo{year}{2023}), \bibinfo{pages}{368--378}.
\newblock


\bibitem[Culver et~al\mbox{.}(2004)]%
        {culver2004exact}
\bibfield{author}{\bibinfo{person}{Tim Culver}, \bibinfo{person}{John Keyser}, {and} \bibinfo{person}{Dinesh Manocha}.} \bibinfo{year}{2004}\natexlab{}.
\newblock \showarticletitle{Exact computation of the medial axis of a polyhedron}.
\newblock \bibinfo{journal}{\emph{Computer Aided Geometric Design}} \bibinfo{volume}{21}, \bibinfo{number}{1} (\bibinfo{year}{2004}), \bibinfo{pages}{65--98}.
\newblock


\bibitem[Dey and Zhao(2002)]%
        {dey2002approximate}
\bibfield{author}{\bibinfo{person}{Tamal~K Dey} {and} \bibinfo{person}{Wulue Zhao}.} \bibinfo{year}{2002}\natexlab{}.
\newblock \showarticletitle{Approximate medial axis as a voronoi subcomplex}. In \bibinfo{booktitle}{\emph{Proceedings of the seventh ACM symposium on Solid modeling and applications}}. \bibinfo{pages}{356--366}.
\newblock


\bibitem[Dey and Zhao(2004)]%
        {dey2004approximating}
\bibfield{author}{\bibinfo{person}{Tamal~K Dey} {and} \bibinfo{person}{Wulue Zhao}.} \bibinfo{year}{2004}\natexlab{}.
\newblock \showarticletitle{Approximating the medial axis from the Voronoi diagram with a convergence guarantee}.
\newblock \bibinfo{journal}{\emph{Algorithmica}} \bibinfo{volume}{38}, \bibinfo{number}{1} (\bibinfo{year}{2004}), \bibinfo{pages}{179--200}.
\newblock


\bibitem[Dou et~al\mbox{.}(2022)]%
        {dou2021coverage}
\bibfield{author}{\bibinfo{person}{Zhiyang Dou}, \bibinfo{person}{Cheng Lin}, \bibinfo{person}{Rui Xu}, \bibinfo{person}{Lei Yang}, \bibinfo{person}{Shiqing Xin}, \bibinfo{person}{Taku Komura}, {and} \bibinfo{person}{Wenping Wang}.} \bibinfo{year}{2022}\natexlab{}.
\newblock \showarticletitle{Coverage Axis: Inner Point Selection for 3D Shape Skeletonization}. In \bibinfo{booktitle}{\emph{Computer Graphics Forum}}, Vol.~\bibinfo{volume}{41}. Wiley Online Library, \bibinfo{pages}{419--432}.
\newblock


\bibitem[Du et~al\mbox{.}(1999)]%
        {du1999centroidal}
\bibfield{author}{\bibinfo{person}{Qiang Du}, \bibinfo{person}{Vance Faber}, {and} \bibinfo{person}{Max Gunzburger}.} \bibinfo{year}{1999}\natexlab{}.
\newblock \showarticletitle{Centroidal Voronoi tessellations: Applications and algorithms}.
\newblock \bibinfo{journal}{\emph{SIAM review}} \bibinfo{volume}{41}, \bibinfo{number}{4} (\bibinfo{year}{1999}), \bibinfo{pages}{637--676}.
\newblock


\bibitem[Foskey et~al\mbox{.}(2003)]%
        {foskey2003efficient}
\bibfield{author}{\bibinfo{person}{Mark Foskey}, \bibinfo{person}{Ming~C Lin}, {and} \bibinfo{person}{Dinesh Manocha}.} \bibinfo{year}{2003}\natexlab{}.
\newblock \showarticletitle{Efficient computation of a simplified medial axis}. In \bibinfo{booktitle}{\emph{Proceedings of the eighth ACM symposium on Solid modeling and applications}}. \bibinfo{pages}{96--107}.
\newblock


\bibitem[Frey and Borouchaki(1999)]%
        {frey1999surface}
\bibfield{author}{\bibinfo{person}{Pascal~J Frey} {and} \bibinfo{person}{Houman Borouchaki}.} \bibinfo{year}{1999}\natexlab{}.
\newblock \showarticletitle{Surface mesh quality evaluation}.
\newblock \bibinfo{journal}{\emph{International journal for numerical methods in engineering}} \bibinfo{volume}{45}, \bibinfo{number}{1} (\bibinfo{year}{1999}), \bibinfo{pages}{101--118}.
\newblock


\bibitem[Ge et~al\mbox{.}(2023)]%
        {ge2023point2mm}
\bibfield{author}{\bibinfo{person}{Mengyuan Ge}, \bibinfo{person}{Junfeng Yao}, \bibinfo{person}{Baorong Yang}, \bibinfo{person}{Ningna Wang}, \bibinfo{person}{Zhonggui Chen}, {and} \bibinfo{person}{Xiaohu Guo}.} \bibinfo{year}{2023}\natexlab{}.
\newblock \showarticletitle{Point2MM: Learning medial mesh from point clouds}.
\newblock \bibinfo{journal}{\emph{Computers \& Graphics}}  \bibinfo{volume}{115} (\bibinfo{year}{2023}), \bibinfo{pages}{511--521}.
\newblock


\bibitem[Giblin and Kimia(2004)]%
        {giblin2004formal}
\bibfield{author}{\bibinfo{person}{Peter Giblin} {and} \bibinfo{person}{Benjamin~B Kimia}.} \bibinfo{year}{2004}\natexlab{}.
\newblock \showarticletitle{A formal classification of 3D medial axis points and their local geometry}.
\newblock \bibinfo{journal}{\emph{IEEE Transactions on Pattern Analysis and Machine Intelligence}} \bibinfo{volume}{26}, \bibinfo{number}{2} (\bibinfo{year}{2004}), \bibinfo{pages}{238--251}.
\newblock


\bibitem[Giesen et~al\mbox{.}(2009)]%
        {giesen2009scale}
\bibfield{author}{\bibinfo{person}{Joachim Giesen}, \bibinfo{person}{Balint Miklos}, \bibinfo{person}{Mark Pauly}, {and} \bibinfo{person}{Camille Wormser}.} \bibinfo{year}{2009}\natexlab{}.
\newblock \showarticletitle{The scale axis transform}. In \bibinfo{booktitle}{\emph{Proceedings of the twenty-fifth annual symposium on Computational geometry}}. \bibinfo{pages}{106--115}.
\newblock


\bibitem[Hesselink and Roerdink(2008)]%
        {hesselink2008euclidean}
\bibfield{author}{\bibinfo{person}{Wim~H Hesselink} {and} \bibinfo{person}{Jos~BTM Roerdink}.} \bibinfo{year}{2008}\natexlab{}.
\newblock \showarticletitle{Euclidean skeletons of digital image and volume data in linear time by the integer medial axis transform}.
\newblock \bibinfo{journal}{\emph{IEEE Transactions on Pattern Analysis and Machine Intelligence}} \bibinfo{volume}{30}, \bibinfo{number}{12} (\bibinfo{year}{2008}), \bibinfo{pages}{2204--2217}.
\newblock


\bibitem[Hu et~al\mbox{.}(2022)]%
        {hu2022immat}
\bibfield{author}{\bibinfo{person}{Jianwei Hu}, \bibinfo{person}{Gang Chen}, \bibinfo{person}{Baorong Yang}, \bibinfo{person}{Ningna Wang}, \bibinfo{person}{Xiaohu Guo}, {and} \bibinfo{person}{Bin Wang}.} \bibinfo{year}{2022}\natexlab{}.
\newblock \showarticletitle{IMMAT: Mesh reconstruction from single view images by medial axis transform prediction}.
\newblock \bibinfo{journal}{\emph{Computer-Aided Design}}  \bibinfo{volume}{150} (\bibinfo{year}{2022}), \bibinfo{pages}{103304}.
\newblock


\bibitem[Hu et~al\mbox{.}(2019)]%
        {Hu2019MATNet}
\bibfield{author}{\bibinfo{person}{Jianwei Hu}, \bibinfo{person}{Bin Wang}, \bibinfo{person}{Lihui Qian}, \bibinfo{person}{Yiling Pan}, \bibinfo{person}{Xiaohu Guo}, \bibinfo{person}{Lingjie Liu}, {and} \bibinfo{person}{Wenping Wang}.} \bibinfo{year}{2019}\natexlab{}.
\newblock \showarticletitle{MAT-Net: Medial Axis Transform Network for 3D Object Recognition}. In \bibinfo{booktitle}{\emph{Proceedings of the 28th International Joint Conference on Artificial Intelligence}} \emph{(\bibinfo{series}{IJCAI'19})}. \bibinfo{pages}{774–781}.
\newblock


\bibitem[Hu et~al\mbox{.}(2023)]%
        {hu2023s3ds}
\bibfield{author}{\bibinfo{person}{Jianwei Hu}, \bibinfo{person}{Ningna Wang}, \bibinfo{person}{Baorong Yang}, \bibinfo{person}{Gang Chen}, \bibinfo{person}{Xiaohu Guo}, {and} \bibinfo{person}{Bin Wang}.} \bibinfo{year}{2023}\natexlab{}.
\newblock \showarticletitle{S3DS: Self-supervised Learning of 3D Skeletons from Single View Images}. In \bibinfo{booktitle}{\emph{Proceedings of the 31st ACM International Conference on Multimedia}}. \bibinfo{pages}{6948--6958}.
\newblock


\bibitem[Hu et~al\mbox{.}(2020)]%
        {hu2020ftetwild}
\bibfield{author}{\bibinfo{person}{Yixin Hu}, \bibinfo{person}{Teseo Schneider}, \bibinfo{person}{Bolun Wang}, \bibinfo{person}{Denis Zorin}, {and} \bibinfo{person}{Daniele Panozzo}.} \bibinfo{year}{2020}\natexlab{}.
\newblock \showarticletitle{Fast tetrahedral meshing in the wild}.
\newblock \bibinfo{journal}{\emph{ACM Transactions on Graphics (TOG)}} \bibinfo{volume}{39}, \bibinfo{number}{4} (\bibinfo{year}{2020}), \bibinfo{pages}{117--1}.
\newblock


\bibitem[Huang et~al\mbox{.}(2024)]%
        {vmas2024}
\bibfield{author}{\bibinfo{person}{Qijia Huang}, \bibinfo{person}{Pierre Kraemer}, \bibinfo{person}{Sylvain Thery}, {and} \bibinfo{person}{Dominique Bechmann}.} \bibinfo{year}{2024}\natexlab{}.
\newblock \showarticletitle{Dynamic Skeletonization via Variational Medial Axis Sampling}. In \bibinfo{booktitle}{\emph{SIGGRAPH Asia 2024 Conference Papers}} (Tokyo, Japan) \emph{(\bibinfo{series}{SA '24})}. \bibinfo{publisher}{Association for Computing Machinery}, \bibinfo{address}{New York, NY, USA}, Article \bibinfo{articleno}{66}, \bibinfo{numpages}{11}~pages.
\newblock
\showISBNx{9798400711312}
\urldef\tempurl%
\url{https://doi.org/10.1145/3680528.3687678}
\showDOI{\tempurl}


\bibitem[Jalba et~al\mbox{.}(2013)]%
        {kustra2013}
\bibfield{author}{\bibinfo{person}{Andrei~C. Jalba}, \bibinfo{person}{Jacek Kustra}, {and} \bibinfo{person}{Alexandru~C. Telea}.} \bibinfo{year}{2013}\natexlab{}.
\newblock \showarticletitle{Surface and Curve Skeletonization of Large 3D Models on the GPU}.
\newblock \bibinfo{journal}{\emph{IEEE Transactions on Pattern Analysis and Machine Intelligence}} \bibinfo{volume}{35}, \bibinfo{number}{6} (\bibinfo{year}{2013}), \bibinfo{pages}{1495--1508}.
\newblock
\urldef\tempurl%
\url{https://doi.org/10.1109/TPAMI.2012.212}
\showDOI{\tempurl}


\bibitem[Jalba et~al\mbox{.}(2015)]%
        {jalba2015unified}
\bibfield{author}{\bibinfo{person}{Andrei~C Jalba}, \bibinfo{person}{Andre Sobiecki}, {and} \bibinfo{person}{Alexandru~C Telea}.} \bibinfo{year}{2015}\natexlab{}.
\newblock \showarticletitle{An unified multiscale framework for planar, surface, and curve skeletonization}.
\newblock \bibinfo{journal}{\emph{IEEE transactions on pattern analysis and machine intelligence}} \bibinfo{volume}{38}, \bibinfo{number}{1} (\bibinfo{year}{2015}), \bibinfo{pages}{30--45}.
\newblock


\bibitem[Koch et~al\mbox{.}(2019)]%
        {koch2019abc}
\bibfield{author}{\bibinfo{person}{Sebastian Koch}, \bibinfo{person}{Albert Matveev}, \bibinfo{person}{Zhongshi Jiang}, \bibinfo{person}{Francis Williams}, \bibinfo{person}{Alexey Artemov}, \bibinfo{person}{Evgeny Burnaev}, \bibinfo{person}{Marc Alexa}, \bibinfo{person}{Denis Zorin}, {and} \bibinfo{person}{Daniele Panozzo}.} \bibinfo{year}{2019}\natexlab{}.
\newblock \showarticletitle{Abc: A big cad model dataset for geometric deep learning}. In \bibinfo{booktitle}{\emph{Proceedings of the IEEE/CVF conference on computer vision and pattern recognition}}. \bibinfo{pages}{9601--9611}.
\newblock


\bibitem[Kong et~al\mbox{.}(2024)]%
        {kong2024quasi}
\bibfield{author}{\bibinfo{person}{Jiayi Kong}, \bibinfo{person}{Chen Zong}, \bibinfo{person}{Jun Luo}, \bibinfo{person}{Shiqing Xin}, \bibinfo{person}{Fei Hou}, \bibinfo{person}{Hanqing Jiang}, \bibinfo{person}{Chen Qian}, {and} \bibinfo{person}{Ying He}.} \bibinfo{year}{2024}\natexlab{}.
\newblock \showarticletitle{Quasi-Medial Distance Field (Q-MDF): A Robust Method for Approximating and Discretizing Neural Medial Axis}.
\newblock \bibinfo{journal}{\emph{arXiv preprint arXiv:2410.17774}} (\bibinfo{year}{2024}).
\newblock


\bibitem[Kustra et~al\mbox{.}(2016)]%
        {kustra2015}
\bibfield{author}{\bibinfo{person}{Jacek Kustra}, \bibinfo{person}{Andrei Jalba}, {and} \bibinfo{person}{Alexandru Telea}.} \bibinfo{year}{2016}\natexlab{}.
\newblock \showarticletitle{Computing refined skeletal features from medial point clouds}.
\newblock \bibinfo{journal}{\emph{Pattern Recognition Letters}}  \bibinfo{volume}{76} (\bibinfo{year}{2016}), \bibinfo{pages}{13--21}.
\newblock
\showISSN{0167-8655}
\urldef\tempurl%
\url{https://doi.org/10.1016/j.patrec.2015.05.007}
\showDOI{\tempurl}
\newblock
\shownote{Special Issue on Skeletonization and its Application}.


\bibitem[Leymarie and Kimia(2007)]%
        {leymarie2007medial}
\bibfield{author}{\bibinfo{person}{Frederic~F Leymarie} {and} \bibinfo{person}{Benjamin~B Kimia}.} \bibinfo{year}{2007}\natexlab{}.
\newblock \showarticletitle{The medial scaffold of 3D unorganized point clouds}.
\newblock \bibinfo{journal}{\emph{IEEE Transactions on Pattern Analysis and Machine Intelligence}} \bibinfo{volume}{29}, \bibinfo{number}{2} (\bibinfo{year}{2007}), \bibinfo{pages}{313--330}.
\newblock


\bibitem[Li et~al\mbox{.}(2015)]%
        {li2015qmat}
\bibfield{author}{\bibinfo{person}{Pan Li}, \bibinfo{person}{Bin Wang}, \bibinfo{person}{Feng Sun}, \bibinfo{person}{Xiaohu Guo}, \bibinfo{person}{Caiming Zhang}, {and} \bibinfo{person}{Wenping Wang}.} \bibinfo{year}{2015}\natexlab{}.
\newblock \showarticletitle{Q-mat: Computing medial axis transform by quadratic error minimization}.
\newblock \bibinfo{journal}{\emph{ACM Transactions on Graphics (TOG)}} \bibinfo{volume}{35}, \bibinfo{number}{1} (\bibinfo{year}{2015}), \bibinfo{pages}{1--16}.
\newblock


\bibitem[Lin et~al\mbox{.}(2021)]%
        {lin2021point2skeleton}
\bibfield{author}{\bibinfo{person}{Cheng Lin}, \bibinfo{person}{Changjian Li}, \bibinfo{person}{Yuan Liu}, \bibinfo{person}{Nenglun Chen}, \bibinfo{person}{Yi-King Choi}, {and} \bibinfo{person}{Wenping Wang}.} \bibinfo{year}{2021}\natexlab{}.
\newblock \showarticletitle{Point2skeleton: Learning skeletal representations from point clouds}. In \bibinfo{booktitle}{\emph{Proceedings of the IEEE/CVF conference on computer vision and pattern recognition}}. \bibinfo{pages}{4277--4286}.
\newblock


\bibitem[Liu and Nocedal(1989)]%
        {liu1989limited}
\bibfield{author}{\bibinfo{person}{Dong~C Liu} {and} \bibinfo{person}{Jorge Nocedal}.} \bibinfo{year}{1989}\natexlab{}.
\newblock \showarticletitle{On the limited memory BFGS method for large scale optimization}.
\newblock \bibinfo{journal}{\emph{Mathematical programming}} \bibinfo{volume}{45}, \bibinfo{number}{1} (\bibinfo{year}{1989}), \bibinfo{pages}{503--528}.
\newblock


\bibitem[Ma et~al\mbox{.}(2012)]%
        {ma20123shrink}
\bibfield{author}{\bibinfo{person}{Jaehwan Ma}, \bibinfo{person}{Sang~Won Bae}, {and} \bibinfo{person}{Sunghee Choi}.} \bibinfo{year}{2012}\natexlab{}.
\newblock \showarticletitle{3D medial axis point approximation using nearest neighbors and the normal field}.
\newblock \bibinfo{journal}{\emph{The Visual Computer}} \bibinfo{volume}{28}, \bibinfo{number}{1} (\bibinfo{year}{2012}), \bibinfo{pages}{7--19}.
\newblock


\bibitem[Miklos et~al\mbox{.}(2010)]%
        {miklos2010sat}
\bibfield{author}{\bibinfo{person}{Balint Miklos}, \bibinfo{person}{Joachim Giesen}, {and} \bibinfo{person}{Mark Pauly}.} \bibinfo{year}{2010}\natexlab{}.
\newblock \showarticletitle{Discrete scale axis representations for 3D geometry}.
\newblock In \bibinfo{booktitle}{\emph{ACM SIGGRAPH 2010 papers}}. \bibinfo{pages}{1--10}.
\newblock


\bibitem[Milenkovic(1993)]%
        {milenkovic1993robust}
\bibfield{author}{\bibinfo{person}{Victor Milenkovic}.} \bibinfo{year}{1993}\natexlab{}.
\newblock \showarticletitle{Robust Construction of the Voronoi Diagram of a Polyhedron.}. In \bibinfo{booktitle}{\emph{CCCG}}, Vol.~\bibinfo{volume}{93}. Citeseer, \bibinfo{pages}{473--478}.
\newblock


\bibitem[Rumpf and Telea(2002)]%
        {rumpf2002continuous}
\bibfield{author}{\bibinfo{person}{Martin Rumpf} {and} \bibinfo{person}{Alexandru Telea}.} \bibinfo{year}{2002}\natexlab{}.
\newblock \showarticletitle{A continuous skeletonization method based on level sets}. In \bibinfo{booktitle}{\emph{Proceedings of the Symposium on Data Visualisation 2002}} (Barcelona, Spain) \emph{(\bibinfo{series}{VISSYM '02})}. \bibinfo{publisher}{Eurographics Association}, \bibinfo{address}{Goslar, DEU}, \bibinfo{pages}{151–ff}.
\newblock
\showISBNx{158113536X}


\bibitem[Saha et~al\mbox{.}(2016)]%
        {saha2016survey}
\bibfield{author}{\bibinfo{person}{Punam~K Saha}, \bibinfo{person}{Gunilla Borgefors}, {and} \bibinfo{person}{Gabriella~Sanniti di Baja}.} \bibinfo{year}{2016}\natexlab{}.
\newblock \showarticletitle{A survey on skeletonization algorithms and their applications}.
\newblock \bibinfo{journal}{\emph{Pattern recognition letters}}  \bibinfo{volume}{76} (\bibinfo{year}{2016}), \bibinfo{pages}{3--12}.
\newblock


\bibitem[Sherbrooke et~al\mbox{.}(1996)]%
        {sherbrooke1996algorithm}
\bibfield{author}{\bibinfo{person}{Evan~C Sherbrooke}, \bibinfo{person}{Nicholas~M Patrikalakis}, {and} \bibinfo{person}{Erik Brisson}.} \bibinfo{year}{1996}\natexlab{}.
\newblock \showarticletitle{An algorithm for the medial axis transform of 3D polyhedral solids}.
\newblock \bibinfo{journal}{\emph{IEEE transactions on visualization and computer graphics}} \bibinfo{volume}{2}, \bibinfo{number}{1} (\bibinfo{year}{1996}), \bibinfo{pages}{44--61}.
\newblock


\bibitem[Siddiqi et~al\mbox{.}(2002)]%
        {siddiqi2002hamilton}
\bibfield{author}{\bibinfo{person}{Kaleem Siddiqi}, \bibinfo{person}{Sylvain Bouix}, \bibinfo{person}{Allen Tannenbaum}, {and} \bibinfo{person}{Steven~W Zucker}.} \bibinfo{year}{2002}\natexlab{}.
\newblock \showarticletitle{Hamilton-jacobi skeletons}.
\newblock \bibinfo{journal}{\emph{International Journal of Computer Vision}}  \bibinfo{volume}{48} (\bibinfo{year}{2002}), \bibinfo{pages}{215--231}.
\newblock


\bibitem[Sobiecki et~al\mbox{.}(2014)]%
        {sobiecki2014comparison}
\bibfield{author}{\bibinfo{person}{Andr{\'e} Sobiecki}, \bibinfo{person}{Andrei Jalba}, {and} \bibinfo{person}{Alexandru Telea}.} \bibinfo{year}{2014}\natexlab{}.
\newblock \showarticletitle{Comparison of curve and surface skeletonization methods for voxel shapes}.
\newblock \bibinfo{journal}{\emph{Pattern Recognition Letters}}  \bibinfo{volume}{47} (\bibinfo{year}{2014}), \bibinfo{pages}{147--156}.
\newblock


\bibitem[Sud et~al\mbox{.}(2005)]%
        {sud2005homotopy}
\bibfield{author}{\bibinfo{person}{Avneesh Sud}, \bibinfo{person}{Mark Foskey}, {and} \bibinfo{person}{Dinesh Manocha}.} \bibinfo{year}{2005}\natexlab{}.
\newblock \showarticletitle{Homotopy-preserving medial axis simplification}. In \bibinfo{booktitle}{\emph{Proceedings of the 2005 ACM symposium on Solid and physical modeling}}. \bibinfo{pages}{39--50}.
\newblock


\bibitem[Tagliasacchi et~al\mbox{.}(2016)]%
        {tagliasacchi20163d}
\bibfield{author}{\bibinfo{person}{Andrea Tagliasacchi}, \bibinfo{person}{Thomas Delame}, \bibinfo{person}{Michela Spagnuolo}, \bibinfo{person}{Nina Amenta}, {and} \bibinfo{person}{Alexandru Telea}.} \bibinfo{year}{2016}\natexlab{}.
\newblock \showarticletitle{3d skeletons: A state-of-the-art report}. In \bibinfo{booktitle}{\emph{Computer Graphics Forum}}, Vol.~\bibinfo{volume}{35}. Wiley Online Library, \bibinfo{pages}{573--597}.
\newblock


\bibitem[Tang et~al\mbox{.}(2025)]%
        {DroneUpdate25}
\bibfield{author}{\bibinfo{person}{Mingfeng Tang}, \bibinfo{person}{Ningna Wang}, \bibinfo{person}{Ziyuan Xie}, \bibinfo{person}{Jianwei Hu}, \bibinfo{person}{Ke Xie}, \bibinfo{person}{Xiaohu Guo}, {and} \bibinfo{person}{Hui Huang}.} \bibinfo{year}{2025}\natexlab{}.
\newblock \showarticletitle{Aerial Path Online Planning for Urban Scene Updation}. In \bibinfo{booktitle}{\emph{ACM SIGGRAPH}}. \bibinfo{pages}{87:1--87:11}.
\newblock


\bibitem[Thiery et~al\mbox{.}(2013)]%
        {thiery2013sphere}
\bibfield{author}{\bibinfo{person}{Jean-Marc Thiery}, \bibinfo{person}{{\'E}milie Guy}, {and} \bibinfo{person}{Tamy Boubekeur}.} \bibinfo{year}{2013}\natexlab{}.
\newblock \showarticletitle{Sphere-meshes: Shape approximation using spherical quadric error metrics}.
\newblock \bibinfo{journal}{\emph{ACM Transactions on Graphics (TOG)}} \bibinfo{volume}{32}, \bibinfo{number}{6} (\bibinfo{year}{2013}), \bibinfo{pages}{1--12}.
\newblock


\bibitem[Wang et~al\mbox{.}(2024b)]%
        {wang2024mattopo}
\bibfield{author}{\bibinfo{person}{Ningna Wang}, \bibinfo{person}{Hui Huang}, \bibinfo{person}{Shibo Song}, \bibinfo{person}{Bin Wang}, \bibinfo{person}{Wenping Wang}, {and} \bibinfo{person}{Xiaohu Guo}.} \bibinfo{year}{2024}\natexlab{b}.
\newblock \showarticletitle{MATTopo: Topology-preserving Medial Axis Transform with Restricted Power Diagram}.
\newblock \bibinfo{journal}{\emph{ACM Transactions on Graphics (TOG)}} \bibinfo{volume}{43}, \bibinfo{number}{4} (\bibinfo{year}{2024}).
\newblock
\urldef\tempurl%
\url{https://doi.org/10.1145/3687763}
\showDOI{\tempurl}


\bibitem[Wang et~al\mbox{.}(2022)]%
        {2022MATFP}
\bibfield{author}{\bibinfo{person}{Ningna Wang}, \bibinfo{person}{Bin Wang}, \bibinfo{person}{Wenping Wang}, {and} \bibinfo{person}{Xiaohu Guo}.} \bibinfo{year}{2022}\natexlab{}.
\newblock \showarticletitle{Computing Medial Axis Transform with Feature Preservation via Restricted Power Diagram}.
\newblock \bibinfo{journal}{\emph{ACM Transactions on Graphics (Proceedings of SIGGRAPH Asia 2022)}} \bibinfo{volume}{41}, \bibinfo{number}{6} (\bibinfo{year}{2022}).
\newblock


\bibitem[Wang et~al\mbox{.}(2024a)]%
        {wang2024coverage}
\bibfield{author}{\bibinfo{person}{Zimeng Wang}, \bibinfo{person}{Zhiyang Dou}, \bibinfo{person}{Rui Xu}, \bibinfo{person}{Cheng Lin}, \bibinfo{person}{Yuan Liu}, \bibinfo{person}{Xiaoxiao Long}, \bibinfo{person}{Shiqing Xin}, \bibinfo{person}{Lingjie Liu}, \bibinfo{person}{Taku Komura}, \bibinfo{person}{Xiaoming Yuan}, {et~al\mbox{.}}} \bibinfo{year}{2024}\natexlab{a}.
\newblock \showarticletitle{Coverage Axis++: Efficient Inner Point Selection for 3D Shape Skeletonization}.
\newblock \bibinfo{journal}{\emph{arXiv preprint arXiv:2401.12946}} (\bibinfo{year}{2024}).
\newblock


\bibitem[Witkin and Heckbert(1994)]%
        {witkin1994using}
\bibfield{author}{\bibinfo{person}{Andrew~P Witkin} {and} \bibinfo{person}{Paul~S Heckbert}.} \bibinfo{year}{1994}\natexlab{}.
\newblock \showarticletitle{Using particles to sample and control implicit surfaces}. In \bibinfo{booktitle}{\emph{Proceedings of the 21st annual conference on Computer graphics and interactive techniques}}. \bibinfo{pages}{269--277}.
\newblock


\bibitem[Xu et~al\mbox{.}(2024)]%
        {xu2024cwf}
\bibfield{author}{\bibinfo{person}{Rui Xu}, \bibinfo{person}{Longdu Liu}, \bibinfo{person}{Ningna Wang}, \bibinfo{person}{Shuangmin Chen}, \bibinfo{person}{Shiqing Xin}, \bibinfo{person}{Xiaohu Guo}, \bibinfo{person}{Zichun Zhong}, \bibinfo{person}{Taku Komura}, \bibinfo{person}{Wenping Wang}, {and} \bibinfo{person}{Changhe Tu}.} \bibinfo{year}{2024}\natexlab{}.
\newblock \showarticletitle{CWF: Consolidating Weak Features in High-quality Mesh Simplification}.
\newblock \bibinfo{journal}{\emph{ACM Transactions on Graphics (TOG)}} \bibinfo{volume}{43}, \bibinfo{number}{4} (\bibinfo{year}{2024}).
\newblock
\showISSN{0730-0301}
\urldef\tempurl%
\url{https://doi.org/10.1145/3658159}
\showDOI{\tempurl}


\bibitem[Xu et~al\mbox{.}(2022)]%
        {xu2022rfeps}
\bibfield{author}{\bibinfo{person}{Rui Xu}, \bibinfo{person}{Zixiong Wang}, \bibinfo{person}{Zhiyang Dou}, \bibinfo{person}{Chen Zong}, \bibinfo{person}{Shiqing Xin}, \bibinfo{person}{Mingyan Jiang}, \bibinfo{person}{Tao Ju}, {and} \bibinfo{person}{Changhe Tu}.} \bibinfo{year}{2022}\natexlab{}.
\newblock \showarticletitle{RFEPS: Reconstructing feature-line equipped polygonal surface}.
\newblock \bibinfo{journal}{\emph{ACM Transactions on Graphics (TOG)}} \bibinfo{volume}{41}, \bibinfo{number}{6} (\bibinfo{year}{2022}), \bibinfo{pages}{1--15}.
\newblock


\bibitem[Yan et~al\mbox{.}(2018)]%
        {yan2018voxel}
\bibfield{author}{\bibinfo{person}{Yajie Yan}, \bibinfo{person}{David Letscher}, {and} \bibinfo{person}{Tao Ju}.} \bibinfo{year}{2018}\natexlab{}.
\newblock \showarticletitle{Voxel cores: Efficient, robust, and provably good approximation of 3d medial axes}.
\newblock \bibinfo{journal}{\emph{ACM Transactions on Graphics (TOG)}} \bibinfo{volume}{37}, \bibinfo{number}{4} (\bibinfo{year}{2018}), \bibinfo{pages}{1--13}.
\newblock


\bibitem[Yan et~al\mbox{.}(2016)]%
        {yan2016erosion}
\bibfield{author}{\bibinfo{person}{Yajie Yan}, \bibinfo{person}{Kyle Sykes}, \bibinfo{person}{Erin Chambers}, \bibinfo{person}{David Letscher}, {and} \bibinfo{person}{Tao Ju}.} \bibinfo{year}{2016}\natexlab{}.
\newblock \showarticletitle{Erosion thickness on medial axes of 3D shapes}.
\newblock \bibinfo{journal}{\emph{ACM Transactions on Graphics (TOG)}} \bibinfo{volume}{35}, \bibinfo{number}{4} (\bibinfo{year}{2016}), \bibinfo{pages}{1--12}.
\newblock


\bibitem[Yang et~al\mbox{.}(2018)]%
        {yang2018dmat}
\bibfield{author}{\bibinfo{person}{Baorong Yang}, \bibinfo{person}{Junfeng Yao}, {and} \bibinfo{person}{Xiaohu Guo}.} \bibinfo{year}{2018}\natexlab{}.
\newblock \showarticletitle{DMAT: Deformable medial axis transform for animated mesh approximation}. In \bibinfo{booktitle}{\emph{Computer Graphics Forum}}, Vol.~\bibinfo{volume}{37}. Wiley Online Library, \bibinfo{pages}{301--311}.
\newblock


\bibitem[Yang et~al\mbox{.}(2020)]%
        {yang2020p2mat}
\bibfield{author}{\bibinfo{person}{Baorong Yang}, \bibinfo{person}{Junfeng Yao}, \bibinfo{person}{Bin Wang}, \bibinfo{person}{Jianwei Hu}, \bibinfo{person}{Yiling Pan}, \bibinfo{person}{Tianxiang Pan}, \bibinfo{person}{Wenping Wang}, {and} \bibinfo{person}{Xiaohu Guo}.} \bibinfo{year}{2020}\natexlab{}.
\newblock \showarticletitle{P2MAT-NET: Learning medial axis transform from sparse point clouds}.
\newblock \bibinfo{journal}{\emph{Computer Aided Geometric Design}}  \bibinfo{volume}{80} (\bibinfo{year}{2020}), \bibinfo{pages}{101874}.
\newblock


\bibitem[Zhong et~al\mbox{.}(2013)]%
        {zhong2013particle}
\bibfield{author}{\bibinfo{person}{Zichun Zhong}, \bibinfo{person}{Xiaohu Guo}, \bibinfo{person}{Wenping Wang}, \bibinfo{person}{Bruno L{\'e}vy}, \bibinfo{person}{Feng Sun}, \bibinfo{person}{Yang Liu}, \bibinfo{person}{Weihua Mao}, {et~al\mbox{.}}} \bibinfo{year}{2013}\natexlab{}.
\newblock \showarticletitle{Particle-based anisotropic surface meshing.}
\newblock \bibinfo{journal}{\emph{ACM Trans. Graph.}} \bibinfo{volume}{32}, \bibinfo{number}{4} (\bibinfo{year}{2013}), \bibinfo{pages}{99--1}.
\newblock


\end{thebibliography}

\includepdf[pages=-]{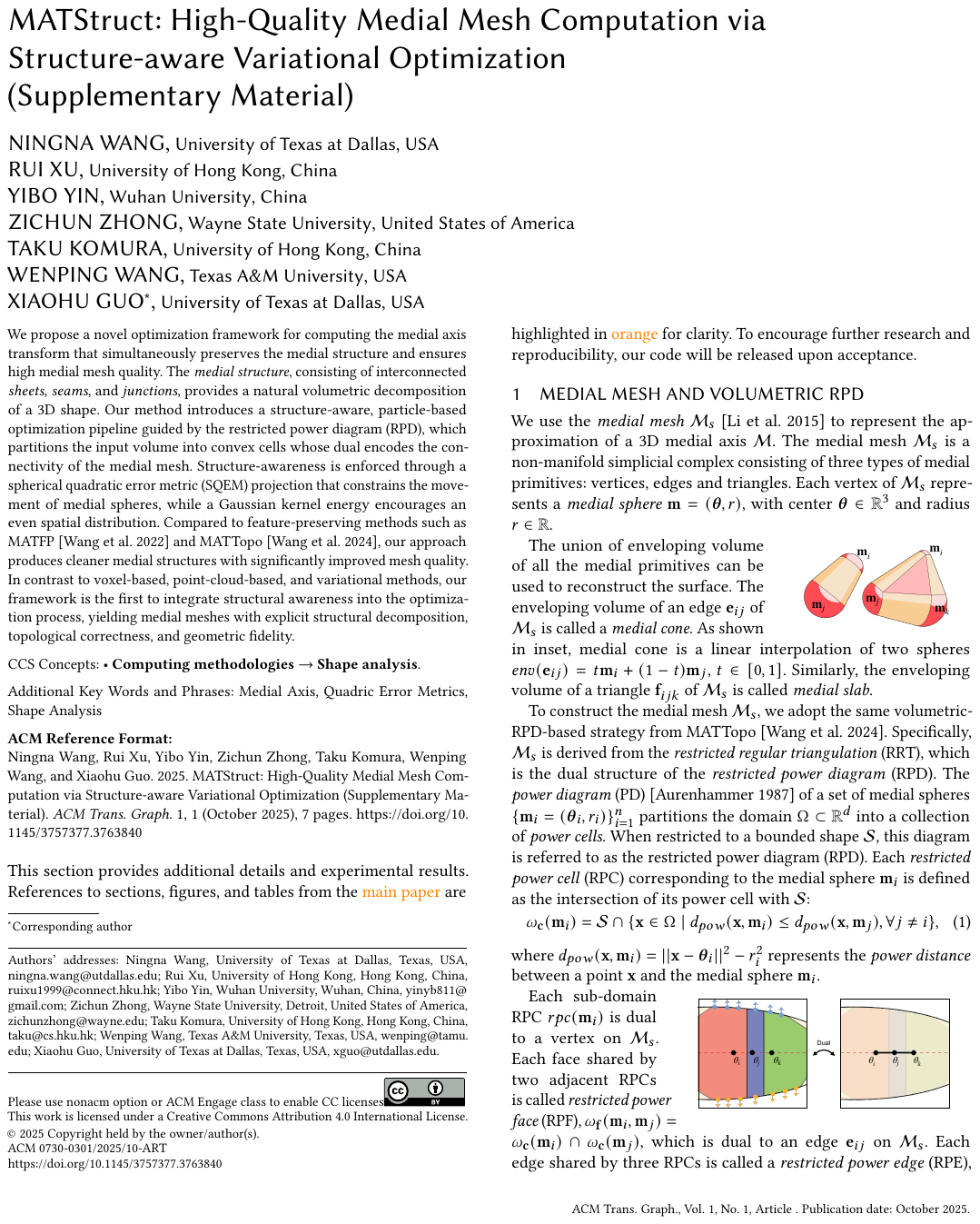}

\end{document}